\documentclass[12pt]{article}
\usepackage{amssymb}

\usepackage{graphics}
\usepackage{epsfig}

\newcommand{\nn}{\nonumber}

\newcommand{\bq}{\begin{eqnarray} }
\newcommand{\eq}{\end{eqnarray} }

\begin{document}

\begin{titlepage}

\begin{flushright}
OSU-HEP-04-1\\
January 2004\\
\end{flushright}
\vspace*{0.5cm}
\begin{center}
{\Large {\bf Constraining $Z^{\prime}$ From Supersymmetry
Breaking} }

\vspace*{1.5cm}
 {\large {\bf O.C. Anoka,$^{a,}$\footnote{E-mail address:anoka@okstate.edu}
 K.S. Babu$^{a,}$\footnote{E-mail address:  babu@okstate.edu}
 and I. Gogoladze$^{a,b,}$\footnote{On a leave of absence from:
Andronikashvili Institute of Physics, GAS, 380077 Tbilisi,
Georgia.  \\ E-mail address: gogoladze.1@nd.edu}}}

 \vspace*{1.cm}
{\it $^a$ Department of Physics, Oklahoma State University\\
Stillwater, OK~74078, USA \\

$^b$ Department of Physics, University of Notre Dame \\Notre Dame,
IN 46556, USA}
\end{center}

 \vspace*{1.5cm}

\begin{abstract}
We suggest and analyze a class of supersymmetric $Z^{\prime}$
models based on the gauge symmetry $U(1)_x=xY-(B-L)$, where $Y$ is
the Standard Model hypercharge. For $1<x<2$, the $U(1)_x$
$D$--term generates positive contributions to the slepton masses,
which is shown to solve the tachyonic slepton problem of anomaly
mediated supersymmetry breaking (AMSB). The resulting models are
very predictive, both in the SUSY breaking sector and in the
$Z^{\prime}$ sector. We find $M_{Z^{\prime}}=2-4$ TeV and the
$Z-Z^{\prime}$ mixing angle $\xi\backsimeq 0.001$. Consistency
with symmetry breaking and AMSB phenomenology renders the
$Z^{\prime}$ ``leptophobic'', with $Br(Z^{\prime}\rightarrow
\ell^+\ell^-)\backsimeq (1-1.6)\%$ and $Br(Z^{\prime}\rightarrow
q\bar{q})\backsimeq 44\%$. The lightest SUSY particle is either
the neutral Wino or the sneutrino in these models.

\end{abstract}

\end{titlepage}

\newpage

\section{Introduction}
One of the simplest extensions of the Standard Model (SM) is
obtained by adding a $U(1)$ factor to the $SU(3)_C\times
SU(2)_L\times U(1)_Y$ gauge structure. Such $U(1)$ factors arise
quite naturally when the SM is embedded in a grand unified group
such as $SO(10)$, $SU(6)$, $E_6$, etc \cite{Hewett,new}. While it
is possible that such $U(1)$ symmetries are broken spontaneously
near the grand unification scale, it is also possible that some of
the $U(1)$ factors survive down to the TeV scale. In fact, if
there is low energy supersymmetry, it is quite plausible that the
$U(1)$ symmetry is broken along with supersymmetry at the TeV
scale. The $Z^{\prime}_ {\chi}$ and $Z^{\prime}_{\psi}$ models
arising from $SO(10)\rightarrow SU(5)\times U(1)_\chi$ and
$E_6\rightarrow SO(10)\times U(1)_\psi$ are two popular extensions
which have attracted much phenomenological attention [1--8].
$Z^{\prime}$ associated with the left--right symmetric extension
of the Standard Model does not require a grand unified symmetry.
Other types of $U(1)$ symmetries, which do not resemble the ones
with a GUT origin, are known to arise in string theory, in the
free--fermionic construction as well as in orbifold and $D$--brane
models [9--11]. Gauge kinetic mixing terms of the type
$B^{\mu\nu}Z^{\prime}_{\mu\nu}$ \cite{Holdom} which will be
generated through renormalization group flow below the unification
scale can further disguise the couplings of the $Z^{\prime}$.

The properties of the $Z^{\prime}$ gauge boson -- its mass, mixing
and couplings to fermions -- associated with the $U(1)$ gauge
symmetry are in general quite arbitrary \cite{babu}. This is
especially so when the low energy theory contains new fermions for
anomaly cancellation. In this paper we propose and analyze a
special class of $U(1)$ models wherein the $Z^{\prime}$ properties
get essentially fixed from constraints of SUSY breaking. We have
in mind the anomaly mediated supersymmetric (AMSB) framework
\cite{Randall,Giudice}. In its minimal version, with the Standard
Model gauge symmetry, it turns out that the sleptons of AMSB
become tachyonic. We suggest the $U(1)$ symmetry, identified as
$U(1)_x=xY-(B-L)$, where $Y$ is the Standard Model hypercharge, as
a solution to the negative slepton mass problem of AMSB. This
symmetry is automatically free of anomalies with the inclusion of
right--handed neutrinos. It is shown that the $D$--term of this
$U(1)_x$ provides positive contributions to the slepton masses,
curing the tachyonic problem . The consistency of symmetry
breaking and the SUSY spectrum points towards a specific set of
parameters in the $Z^{\prime}$ sector. For example, $1<x<2$ is
needed for the positivity of the left--handed and the right-handed
slepton masses. Furthermore, the $U(1)_x$ gauge coupling, $g_x$,
is fixed to be between 0.4--0.5. The resulting $Z^{\prime}$ is
found to be ``leptophobic'' \cite{leptophobic} with
$Br(Z\rightarrow \ell^+\ell^-)\simeq (1-1.6)\%$ and
$Br(Z\rightarrow q\bar{q})\simeq 44\%$.

AMSB models are quite predictive as regards the SUSY spectrum.
 The masses of the scalar components of the chiral
supermultiplets  in AMSB scenario are given by
\cite{Randall,Giudice}
\begin{eqnarray}\label{A1}
(m^2)^{\phi_{j}}_{\phi_{i}}=\frac{1}{2}M_{aux}^2\left[\beta(Y)\frac{\partial}{\partial{Y}}\gamma^{\phi_{j}}_{\phi_{i}}
+\beta(g)\frac{\partial}{\partial{g}}\gamma^{\phi_{j}}_{\phi_{i}}\right],
\end{eqnarray}
where summations over the gauge couplings $g$ and the Yukawa
couplings $Y$ are assumed. $\gamma^{\phi_{j}}_{\phi_{i}}$ are the
one--loop anomalous dimensions, $\beta (Y)$ is the beta function
for the Yukawa coupling $Y$, and $\beta (g)$ is the beta function
for the gauge coupling $g$. $M_{aux}$ is the vacuum expectation
value of a ``compensator superfield" \cite{Randall} which sets the
scale of SUSY breaking. The gaugino mass $M_{g}$, the trilinear
soft supersymmetry breaking term $A_Y$ and the bilinear SUSY
breaking term $B$ are given by \cite{Randall,Giudice}
\begin{eqnarray}\label{A2}
M_{g}=\frac{\beta{(g)}}{g}M_{aux},\,\,\,\,\,A_{Y}=-\frac{\beta{(Y)}}{Y}M_{aux},\,\,\,\,\,B=-M_{aux}(\gamma_{H_u}+\gamma_{H_d}).
\end{eqnarray}
We see that the SUSY masses are completely fixed in the AMSB
framework once the spectrum of the theory and $M_{aux}$ are
specified.

 The negative slepton
mass problem arises in AMSB because in Eq. (1) the gauge beta
functions for $SU(2)_L$ and $U(1)_Y$ are positive,
$\gamma_{\phi_i}^{\phi_j}$ are negative, and the Yukawa couplings
are small for the first two families of sleptons. In our
$Z^{\prime}$ models, there are additional positive contributions
from the $U(1)_x$ $D$--terms which render these masses positive.

In Ref. \cite{JJ} the negative slepton mass problem of AMSB has
been solved with explicit Fayet--Iliopoulos terms added to the
theory. In contrast, in our models, the $D$--term is calculable,
which makes the $Z^{\prime}$ sector more predictive. We find
$M_{Z^{\prime}}=2-4$ TeV and the $Z-Z^{\prime}$ mixing angle
$\xi\simeq 0.001$. Constraints from the electroweak precision
observables are satisfied, with the $Z^{\prime}$ model giving a
slightly better fit compared to the Standard Model.

Other attempts to solve the negative slepton mass problem of AMSB
generally assume TeV--scale new physics [18--20] or a universal
scalar mass of non--AMSB origin \cite{Fengm}. In Ref. \cite{anoka}
we have shown how a non--Abelian horizontal symmetry which is
asymptotically free solves the problem. Some of the techniques we
use here for the symmetry breaking analysis are similar to Ref.
\cite{anoka}.

The plan of the paper is as follows. In section 2 we introduce our
model. In section 3 we analyze the Higgs potential of the model.
In section 4 we present formulas for the SUSY spectrum. Section 5
contains our numerical results for the SUSY spectrum as well as
for the $Z^{\prime}$ mass and mixing. In section 6 we analyze the
partial decay modes of the $Z^{\prime}$.
  In section 7 we analyze other experimental test of the model.
  Here we show the consistency of our models with the precision
  electroweak data.
 Section 8 has our conclusions. In an Appendix we give the relevant expressions for
the beta functions, anomalous dimensions as well as for the soft
masses.

\section{$U(1)_{x}$ Model}

We present our model in this section. We consider adding an extra
$U(1)$ gauge group to the Standard Model gauge structure of MSSM.
The model is then based on the gauge group $SU(3)_{C}\otimes
SU(2)_{L}\otimes U(1)_{Y}\otimes U(1)_{x}$, where the $U(1)_x$
charge is given by the following linear combination of hypercharge
$Y$ and $B-L$:
\begin{eqnarray}\label{A5}
U(1)_{x}&=&x Y-(B-L).
\end{eqnarray}
The particle content of the model and the $U(1)_x$ charge
assignment are shown in Table 1. Besides the MSSM particles, the
model has new particles $
\{\nu_i^c,\,\,\,\nu^c,\,\,\,\bar{\nu}^c,\,\,\, S_+
\,\,\,$and$\,\,\, S_- \}$ which are all singlets of the Standard
Model gauge group.

\begin{table}[h]
\begin{center}\small{
\begin{tabular}{|c|c|c|c|c|c|c|c|c|c|c|c|c|}\hline
\rule[5mm]{0mm}{0pt}
Superfield&$Q_{i}$&$u^c_i$&$d^c_i$&$L_{i}$&$e^{c}_{i}$&$H_{u}$&$H_{d}$&$\nu^{c}_{i}$&$\nu^{c}$&$\bar{\nu}^{c}$&$S_{+}$&$S_{-}$
\\\hline$U(1)_{x}$&$\,\,\,\,\frac{x}{6}-\frac{1}{3}$&$-\frac{2x}{3}+\frac{1}{3}$&$\,\,\,\,\frac{x}{3}+\frac{1}{3}$
&$-\frac{x}{2}+1$&$x-1$&$\frac{x}{2}$&$-\frac{x}{2}$&$-1$&$-1$&$1$&$2$&
$-2$\\\hline \end{tabular} \caption{\footnotesize Particle content
and charge assignment of the $U(1)_x$ model. Here $i $ = $1-3$ is
the family index.}
  \label{D0}}
\end{center}
\end{table}
In order for $\tilde{L}_i$ and $\tilde{e}^c_i$ sleptons to have
positive mass--squared from the $U(1)_x$ $D$--term, the charges of
$L_i$ and $e^c_i$ must be of the same sign. This is possible only
for $1<x<2$. We shall confine to this range of $x$, which is an
important restriction on this class of models. The $\nu^c_i$
fields are needed for $U(1)_x$ anomaly cancellation. $S_+$ and
$S_-$ are the Higgs superfields responsible for $U(1)_x$ symmetry
breaking. The $\nu^c+\bar{\nu}^c$ pair facilitates symmetry
breaking within the AMSB framework. The superpotential of the
model consistent with the gauge symmetries is given by:
\begin{eqnarray}\label{A6}
W&=&\left(Y_{u}\right)_{ij}Q_{i}H_{u}u_{j}^{c}
+\left(Y_{d}\right)_{ij}Q_{i}H_{d}d_{j}^{c}+\left(Y_{l}\right)_{ij}L_{i}H_{d}e_{j}^{c}+\mu
H_{u}H_{d}\nonumber\\&+&\mu^{\prime} S_{+}S_{-}+
\sum_{i=1}^{3}f_{\nu_{i}^c}\nu_{i}^c\nu_{i}^c
S_{+}+f_{\nu^c}\nu^c\nu^c S_{+}+h\bar{\nu}^c\bar{\nu}^c
S_{-}+M_{\nu^c}\nu^c\bar{\nu}^c.
\end{eqnarray}
Here $i,j=1,2,3$ are the family indices. The mass parameters $\mu$
and $\mu^{\prime}$ are of order TeV, which may have a natural
origin in AMSB \cite{Randall}. In general, one can write
additional mass terms of the form $M_i\nu^c_i\bar{\nu}^c$ in the
superpotential. Such terms will have very little effect on the
symmetry breaking analysis that follows. We forbid such mass terms
by invoking a discrete symmetry (such as a $Z_2$) which
differentiates $\nu^c$ from $\nu_i^c$.

Small neutrino masses are induced in the model through the seesaw
mechanism. However, the $\nu_i^c$ fields, which remain light to
the TeV scale, are not to be identified as the traditional
right--handed neutrinos involved in the seesaw mechanism. The
heavy fields which are integrated out have $U(1)_x$--invariant
mass terms. Specifically, the following
 effective nonrenormalizable operators emerge after
integrating out the heavy neutral lepton  fields:
\begin{eqnarray}\label{A7}
L^{\nu}_{eff}=\frac{Y^2_{\nu_{ij}}}{M_N^2}L_iL_jH_uH_uS_-.
\end{eqnarray}
Here $M_N$ represents the masses of the heavy neutral leptons. For
$M_N \sim 10^9$ GeV and $\langle S_-\rangle \sim$ TeV, sub--eV
neutrino masses are obtained. Note that we have not allowed
neutrino Dirac Yukawa couplings of the form
$h_{\nu_{ij}}L_i\nu^c_jH_u$, which would generate Majorana masses
of order MeV for the light neutrinos. We forbid such terms by a
global symmetry $G$, either discrete or continuous. In our
numerical examples we shall assume this symmetry to be
non--Abelian, with $\nu_i^c$ transforming as a triplet [for
example, $G$ can be $O(3)$, $S_4$, $A_4$, etc.]. Such a symmetry
would imply that $f_{\nu_i^c}$ in Eq. (\ref{A6}) are equal for
$i=1-3$.

\section{Symmetry Breaking}
The scalar potential (involving $H_u,\,H_d,\,S_+,\,S_-$ fields) of
the model is given by:
\begin{eqnarray}\label{A8}
V&=&(M^2_{H_u}+\mu^2)|H_u|^2+(M^2_{H_d}+\mu^2)|H_d|^2+(M^2_{S_+}+\mu^{\prime
2})|S_+|^2+(M^2_{S_-}+\mu^{\prime 2})|S_-|^2\nonumber\\&+&
B\mu(H_uH_d+h.c.)+B^{\prime}\mu^{\prime}(S_+S_-+h.c.)+\frac{1}{8}(g_{1}^2+g_{2}^2)(|H_u|^2-|H_d|^2)^2\nonumber\\
&+&
\frac{1}{2}g_{2}^2|H_{u}H_d|^2+\frac{1}{2}g_x^2\left(\frac{x}{2}|H_u|^2-\frac{x}{2}|H_d|^2+2|S_+|^2-2|S_-|^2\right)^2,
\end{eqnarray}
where the last term is the $U(1)_x$ $D$ term. The $B$ and the
$B^{\prime}$ terms for the model are given by
\begin{eqnarray}\label{A91}
B=-(\gamma_{H_u}+\gamma_{H_d})M_{aux}\,\,\,\,and\,\,\,\,\,B^{\prime}=-(\gamma_{S_+}+\gamma_{S_-})M_{aux},
\end{eqnarray}
where the $\gamma$'s are the one--loop anomalous dimensions given
in the Appendix, Eqs. (115)--(116), (120)--(121).

 We parameterize the VEVs of $H_u,\,H_d,\,S_+\,$and$ \,\,S_-$ as
\begin{eqnarray}\label{A9}
\langle H_u\rangle=\pmatrix{0\cr \upsilon_u},\,\,\,\,\langle
H_d\rangle=\pmatrix{\upsilon_d\cr0},\,\,\,\,\langle
S_+\rangle=z,\,\,\,\,\langle S_-\rangle=y.
\end{eqnarray}
 In minimizing
the potential, we have to keep in mind the fact that the VEVs of
$\langle S_+\rangle$ and $\langle S_-\rangle$ should be much
larger than the VEVs of $\langle H_u\rangle$ and $\langle
H_d\rangle$ for a consistent picture. In addition, the VEV of
$\langle S_+\rangle$ should be greater than the VEV of $\langle
S_-\rangle$ in order for the $D$--term contribution to the slepton
masses to be positive. We have checked explicitly that all the
above--mentioned conditions are satisfied at the local minimum for
a restricted choice of model parameters. The physical Higgs bosons
as well as the sleptons acquire positive mass--squared, while
generating a $Z^{\prime}$ mass and $Z-Z^{\prime}$ mixing angle
consistent with experimental constraints.

Minimization of the potential leads to the following conditions:
\begin{eqnarray}\label{A11}
\sin{2\beta}&=&\frac{2B\mu}{2\mu^2+M_{H_u}^2+M_{H_d}^2},\\
\frac{M_Z^2}{2}&=&-\mu^2+\frac{M_{H_d}^2-M_{H_u}^2\tan^2{\beta}}{\tan^{2}{\beta}-1}-\frac{x^2g_x^2\upsilon^2}{4}-\frac{xg_x^2u^2\cos{2\psi}}{\cos{2\beta}},\\
\sin{2\psi}&=&\frac{-2B^{\prime}\mu^{\prime}}{2\mu^{\prime 2}+M_{S_+}^2+M_{S_-}^2},\\
\frac{M_{Z^\prime}^2}{2}&=&-\mu^{\prime
2}+\frac{M_{S_-}^2-M_{S_+}^2\tan^2{\psi}}{(\tan^2{\psi}-1)}+\frac{x^2g_{x}^2\upsilon^2}{4}-\frac{xg_x^2\upsilon^2\cos{2\beta}}{\cos{2\psi}}.
\end{eqnarray}
Here $M_{Z^{\prime}}^2=\frac{x^2g_x^2\upsilon^2}{2}+8g_x^2u^2$,
$\tan{\beta}=\frac{\upsilon_u}{\upsilon_d}$,
$\tan{\psi}=\frac{z}{y}$,
$\sqrt{\upsilon_u^2+\upsilon_d^2}=\upsilon=174 $ GeV and
$\sqrt{z^2+y^2}=u$.

To see the consistency of symmetry breaking, we need to calculate
the Higgs boson mass--squared and establish that they are all
positive. We parameterize the Higgs fields (in the unitary gauge)
as
\begin{eqnarray}\label{A10}
&&H_u=\pmatrix{H^+\sin{\beta}\cr
\upsilon_u+\frac{1}{\sqrt{2}}(\phi_2+i\cos{\beta}\,\phi_3)},\,\,\,\,\langle
H_d\rangle=\pmatrix{\upsilon_d+\frac{1}{\sqrt{2}}(\phi_1+i\sin{\beta}\,\phi_3)\cr
H^-\cos{\beta}},\nonumber\\
&&S_+=z+\frac{1}{\sqrt{2}}(\phi_4+i\cos{\psi}\,\phi_5),\,\,\,\,S_-=y+\frac{1}{\sqrt{2}}(\phi_6+i\sin{\psi}\,\phi_5).
\end{eqnarray}

The CP--odd Higgs bosons $\{\phi_3,\,\,\,\phi_5\}$ have masses
given by
\begin{eqnarray}\label{A14}
m^2_A&=&\frac{2B\mu}{\sin{2\beta}},\,\,\,\,\,
m^2_{A^{\prime}}=-\frac{2B^{\prime}\mu^{\prime}}{\sin{2\psi}}.
\end{eqnarray}

 The mass
matrix for the CP--even neutral Higgs bosons
$\{\phi_1,\,\,\phi_2,\,\,\phi_4,\,\,\phi_6\}$ is given by
\begin{eqnarray}\label{A12}
(\mathcal{M}^2)_{cp-even}&=&\pmatrix{(\mathcal{M}^2)_{11}&(\mathcal{M}^2)_{12}&-2xg_x^2\upsilon_d
z&2xg_x^2\upsilon_d y\cr
(\mathcal{M}^2)_{12}&(\mathcal{M}^2)_{22}&2xg_x^2\upsilon_u
z&-2xg_x^2\upsilon_u y\cr -2xg_x^2\upsilon_d z&2xg_x^2\upsilon_u
z&(\mathcal{M}^2)_{33}&(\mathcal{M}^2)_{34}\cr 2xg_x^2\upsilon_d
y&-2xg_x^2\upsilon_u y&(\mathcal{M}^2)_{34}&(\mathcal{M}^2)_{44}},
\end{eqnarray}
where
\begin{eqnarray}\label{A13}
(\mathcal{M}^2)_{11}&=&\,\,\,\,\,m^2_A\sin^2{\beta}+M_Z^2\cos^2{\beta}+\frac{1}{2}(x^2g_x^2\upsilon^2\cos^2{\beta}),\\
(\mathcal{M}^2)_{12}&=&-m^2_A\sin{\beta}\cos{\beta}-M_Z^2\sin{\beta}\cos{\beta}-\frac{1}{2}x^2g_x^2\upsilon^2\sin{\beta}\cos{\beta},\\
(\mathcal{M}^2)_{22}&=&\,\,\,\,\,m^2_A\cos^2{\beta}+M_Z^2\sin^2{\beta}+\frac{1}{2}(x^2g_x^2\upsilon^2\sin^2{\beta}),\\
(\mathcal{M}^2)_{33}&=&\,\,\,\,\,m^2_{A^{\prime}}\cos^2{\psi}+8g_x^2z^2,\\
(\mathcal{M}^2)_{34}&=&-m^2_{A^{\prime}}\sin{\psi}\cos{\psi}-8g_x^2yz,\\
(\mathcal{M}^2)_{44}&=&\,\,\,\,\,m^2_{A^{\prime}}\sin^2{\psi}+8g_x^2y^2.
\end{eqnarray}
It is instructive to analyze the effect of the $U(1)_x$ $D$--term
on the mass of the lightest MSSM Higgs boson $h$. Consider the
upper left $2\times 2$ sub sector of the CP--even Higgs boson mass
matrix. It has eigenvalues given by
\begin{eqnarray}\label{A150}
\lambda_{1,2}&=&\frac{1}{2}\left[m^2_A+M_Z^2+\frac{x^2g_x^2\upsilon^2}{2}\right.\nonumber\\&\mp&\left.\sqrt{\left(m^2_A+M_Z^2+\frac{x^2g_x^2\upsilon^2}{2}\right)^2-4m^2_AM_Z^2\cos^2{2\beta}-4m^2_A\left(\frac{x^2g_x^2\upsilon^2}{2}\right)\cos^2{2\beta}}\right]
\end{eqnarray}
From Eq. (\ref{A150}) we obtain an upper limit on $m_{h}$:
\begin{eqnarray}\label{A151}
m_{h}\leqslant
\sqrt{\frac{x^2g_x^2\upsilon^2}{2}+M_Z^2}|\cos{2\beta}|.
\end{eqnarray}
The mixing between the doublets and the singlets will reduce the
upper limit further. In fact, we find this mixing effect to be
significant.

The lower $2\times 2$ subsector of Eq. (\ref{A12}) has eigenvalues
\begin{eqnarray}\label{A154}
\lambda^{\prime}_{1,2}=\frac{1}{2}\left[8g_x^2u^2+m^2_{A^{\prime}}\mp\sqrt{(8g_x^2u^2+m^2_{A^{\prime}})^2-4m^2_{A^{\prime}}(8g_x^2u^2)\cos^2{2\psi}}\right].
\end{eqnarray}
From Eq. (\ref{A154})  we obtain an upper bound of the lightest
Higgs mass for the $SU$(2) singlet sector:
\begin{eqnarray}\label{A155}
m_{h^{\prime }}\leqslant m_{A^{\prime}}|\cos{2\psi}|.
\end{eqnarray}
The above upper limit on $m_{h^{\prime }}$ is affected only
minimally by the mixing between the doublet and the singlet Higgs
fields.

As in the MSSM, the mass of the charged Higgs boson $H^{\pm}$ is
given by
\begin{eqnarray}\label{A152}
m^2_{H^\pm}&=&m^2_A+M_W^2.
\end{eqnarray}

We now turn to the supersymmetric fermion masses. The (Majorana)
mass matrix of the  neutralinos
$\{\tilde{B},\,\tilde{W_3},\,\tilde{H_d^0},\,\tilde{H_u^0},\,\tilde{B^{\prime}},\,\tilde{S_+},\,\tilde{S_-}\}$
is given by
\begin{eqnarray}\label{A16}
\mathcal{M}^{(0)}=\pmatrix{M_1&0&-\frac{\upsilon_d}{\sqrt{2}}g_1&\frac{\upsilon_u}{\sqrt{2}}g_1&0&0&0\cr
0&M_2&\frac{\upsilon_d}{\sqrt{2}}g_2&-\frac{\upsilon_u}{\sqrt{2}}g_2&0&0&0\cr
-\frac{\upsilon_d}{\sqrt{2}}g_1&\frac{\upsilon_d}{\sqrt{2}}g_2&0&-\mu&-\frac{\upsilon_d}{\sqrt{2}}xg_x&0&0\cr
\frac{\upsilon_u}{\sqrt{2}}g_1
&-\frac{\upsilon_u}{\sqrt{2}}g_2&-\mu&0&\frac{\upsilon_u}{\sqrt{2}}xg_x&0&0\cr
0&0&-\frac{\upsilon_d}{\sqrt{2}}xg_x
&\frac{\upsilon_u}{\sqrt{2}}xg_x&M^{\prime}_1&2\sqrt{2}g_{x}z&-2\sqrt{2}g_{x}y\cr
0&0&0&0&2\sqrt{2}g_xz&0&\mu^{\prime}\cr
0&0&0&0&-2\sqrt{2}g_{x}y&\mu^{\prime}&0},
\end{eqnarray}
where $M_1,\,M_1^{\prime}\,$and$\,M_2$ are the gaugino masses for
$U(1)_Y,\,U(1)_x \,$and$\,SU(2)_L$. The physical neutralino masses
$m_{\tilde{\chi}_i^0}$ ($i=$1--7) are obtained as the eigenvalues
of this mass matrix. We denote the diagonalizing matrix as $O$:
\begin{eqnarray}\label{A161}
O\mathcal{M}^{(0)}O^T=diag\{m_{\tilde{\chi}_1^0},\,\,m_{\tilde{\chi}_2^0},\,\,m_{\tilde{\chi}_3^0},\,\,m_{\tilde{\chi}_4^0},\,\,m_{\tilde{\chi}_5^0},\,\,m_{\tilde{\chi}_6^0},\,\,m_{\tilde{\chi}_7^0}\}.
\end{eqnarray}

In the basis $\{\tilde{W}^+,\,\tilde{H}_u^+\}$,
$\{\tilde{W}^-,\,\tilde{H}_d^-\}$ the chargino (Dirac) mass matrix
is
\begin{eqnarray}\label{A17}
\mathcal{M}^{(c)}=\pmatrix{M_2&g_2\upsilon_d\cr
g_2\upsilon_{u}&\mu}.
\end{eqnarray}
This matrix is diagonalized by a biunitary transformation
$V^{\ast}\mathcal{M}^{(c)}U^{-1}=diag\{m_{\tilde{\chi}_1^{\pm}},\,\,m_{\tilde{\chi}_2^{\pm}}\}$.

The $Z-Z^{\prime}$ mixing matrix is given by
 \begin{eqnarray}\label{A24}
 \mathcal{M}_{Z-Z^{\prime}}^2=\pmatrix{M_Z^2&\gamma M_Z^2\cr \gamma
 M_Z^2&M_{Z^{\prime}}^2},
 \end{eqnarray}
 where
\begin{eqnarray}\label{A25}
\gamma=\frac{-xg_x}{\sqrt{g_1^2+g_2^2}},\,\,\,\,M_Z^2=\frac{\upsilon^2}{2}(g_1^2+g_2^2)
,\,\,\,\,M_{Z^{\prime}}^2=\frac{x^2g_{x}^2\upsilon^2}{2}+8g_{x}^2u^2.
 \end{eqnarray}
 The physical mass eigenstates $Z_1$ and $Z_2$ with masses $M_{Z_1}$, $M_{Z_2}$ are
\begin{eqnarray}\label{A26}
Z_1&=&Z\cos{\xi}+Z^{\prime}\sin{\xi},\\
Z_2&=&-Z\sin{\xi}+Z^{\prime}\cos{\xi},
\end{eqnarray}
where
\begin{eqnarray}\label{A28}
M^2_{Z_{1},Z_{2}}=\frac{1}{2}\left[M_Z^2+M_{Z^{\prime}}^2\mp
\sqrt{(M_Z^2-M_{Z^{\prime}}^2)^2+4\gamma^2M_Z^4}\right].
\end{eqnarray}
The $Z-Z^{\prime}$ mixing angle $\xi$ is given by
\begin{eqnarray}\label{A27}
\xi=\frac{1}{2}\arctan{\left(\frac{2\gamma
M_Z^2}{M_{Z}^2-M_{Z^{\prime}}^2}\right)}\simeq -\gamma
M_{Z}^2/M_{Z^{\prime}}^2.
\end{eqnarray}
We have ignored kinetic mixing of the form
$B^{\mu\nu}Z^{\prime}_{\mu\nu}$ in the Lagrangian
\cite{Holdom,babu}.

The masses of the heavy right--handed neutrinos are given by
\begin{eqnarray}\label{A381}
m_{\nu^c_i}=f_{\nu_{i}^c}z,
\end{eqnarray}
where $i=1-3$ is the family index. The fourth right--handed
neutrino $\nu^c$ mixes with the $\bar{\nu}^c$ field forming two
Majorana fermions. The masses are the eigenvalues of the mass
matrix
\begin{eqnarray}\label{A382}
M_{\nu^c\bar{\nu}^c}=\pmatrix{f_{\nu^c}z&M_{\nu^c}\cr M_{\nu^c}&h
y},
\end{eqnarray}
where $M_{\nu^c}$ is the mass parameter that appears in the
superpotential of Eq. (\ref{A6}). We denote the eigenstates of
this matrix as $\omega_1,\,\omega_2$ and the mass eigenvalues as
$m_{\omega_1}$ and $m_{\omega_2}$.

\section{The SUSY Spectrum}

\subsection{Slepton masses}
 The slepton mass--squareds are given by the
eigenvalues of the mass matrices
\begin{eqnarray}\label{A29}
M^2_{\tilde{l}} =\pmatrix{m^2_{\tilde{l}_i} &
m_{e_{i}}\left(A_{Y_{l_{i}}}-\mu\tan{\beta}\right)\cr
m_{e_{i}}\left(A_{Y_{l_{i}}}-\mu\tan{\beta}\right)&
m^2_{\tilde{e}^{c}_i}},
\end{eqnarray}
where $i=e,\,\mu,\,\tau$, and
\begin{eqnarray}\label{A30}
m^2_{\tilde{l}_i}&=&\frac{M_{aux}^2}{(16\pi^2)}\left[Y_{l_{i}}\beta(Y_{l_{i}})-\left(\frac{3}{2}g_{2}\beta(g_{2})
+\frac{3}{10}g_{1}\beta(g_{1})+2\left(1-\frac{x}{2}\right)^2g_{x}\beta(g_{x})\right)\right]\nonumber\\&+&m_{e_{i}}^2+\left(-\frac{1}{2}+\sin^2{\theta_{W}}\right)\cos{2\beta}M_{Z}^{2}+2g_{x}^2\left(1-\frac{x}{2}\right)(z^2-y^2),\\
m^2_{\tilde{e}^{c}_i}&=&\frac{M_{aux}^2}{(16\pi^2)}\left[2Y_{l_{i}}\beta(Y_{l_{i}})-\left(\frac{6}{5}g_{1}\beta(g_{1})+2(x-1)^2g_{x}\beta(g_{x})\right)\right]\nonumber\\
&+&m_{e_{i}}^2-\sin^2{\theta_{W}}\cos{2\beta}M_{Z}^{2}+2g_{x}^2(x-1)(z^2-y^2).
\end{eqnarray}
The SUSY soft masses are calculated from the RGE given in the
Appendix [Eqs. (124), (\ref{A199})]. Note the positive
contribution from the $U(1)_{x}$ $D$--terms in Eqs.
(\ref{A30})--(40), given by the terms
$+2g_{x}^2(1-\frac{x}{2})(z^2-y^2)$ and $+2g_{x}^2(x-1)(z^2-y^2)$.
There are also negative contributions proportional to
$\beta\,(g_x)$, but in our numerical solutions, the positive
$D$--term contributions are larger than the negative
contributions. We seek solutions where $z=\langle S_+\rangle$ and
$y=\langle S_-\rangle$ are much larger than $\upsilon_u$,
$\upsilon_d$, of order TeV, with $z\gtrsim y$.

The left--handed sneutrino masses are given by
\begin{eqnarray}\label{A36}
m^2_{\tilde{\nu}_{L_i}}&=&\frac{M_{aux}^2}{(16\pi^2)}\left[-\frac{3}{2}g_{2}\beta(g_{2})
-\frac{3}{10}g_{1}\beta(g_{1})-2\left(1-\frac{x}{2}\right)^2g_{x}\beta(g_{x})\right]\nonumber\\&+&\frac{1}{2}\cos{2\beta}M_{Z}^{2}+2g_{x}^2\left(1-\frac{x}{2}\right)(z^2-y^2).
\end{eqnarray}
\subsection{Squark  masses}
The mixing matrix for the squark sector is similar to the slepton
sector. The diagonal entries of the up and the down squark mass
matrices are given by
\begin{eqnarray}\label{A32}
m_{\tilde{U}_{i}}^2&=&(m^{2}_{soft})_{\tilde{Q}_{i}}^{\tilde{Q}_{i}}+m_{U_{i}}^{2}+\frac{1}{6}\left(4M_{W}^{2}-M_{Z}^{2}\right)\cos{2\beta}+2g_{x}^2\left(\frac{x}{6}-\frac{1}{3}\right)(z^2-y^2),\nonumber\\
m_{\tilde{U}^{c}_{i}}^2&=&(m^{2}_{soft})_{\tilde{U}_{i}^{c}}^{\tilde{U}_{i}^{c}}+m_{U_{i}}^{2}-\frac{2}{3}\left(M_{W}^{2}-M_{Z}^{2}\right)\cos{2\beta}+2g_{x}^2\left(-\frac{2x}{3}+\frac{1}{3}\right)(z^2-y^2),\nonumber\\
m_{\tilde{D}_{i}}^2&=&(m^{2}_{soft})_{\tilde{Q}_{i}}^{\tilde{Q}_{i}}+m_{D_{i}}^{2}-\frac{1}{6}\left(2M_{W}^{2}+M_{Z}^{2}\right)\cos{2\beta}+2g_{x}^2\left(\frac{x}{6}-\frac{1}{3}\right)(z^2-y^2),\nonumber\\
m_{\tilde{D}_{i}^{c}}^2&=&(m^{2}_{soft})_{\tilde{D}_{i}^{c}}^{\tilde{D}_{i}^{c}}+
m_{D_{i}}^{2}+\frac{1}{3}\left(M_{W}^{2}-M_{Z}^{2}\right)\cos{2\beta}+2g_{x}^2\left(\frac{x}{3}+\frac{1}{3}\right)(z^2-y^2).
\end{eqnarray}
Here $m_{U_{i}}$ and $m_{D_{i}}$ are quark masses of different
generations, $i$ = 1, 2, 3. The squark soft masses are obtained
from the RGE as
\begin{eqnarray}\label{A.33}
(m^{2}_{soft})_{\tilde{Q}_{i}}^{\tilde{Q}_{i}}&=&\frac{M_{aux}^{2}}{16\pi^2}\left[Y_{u_{i}}
\beta{(Y_{u_{i}})}+Y_{d_{i}}\beta{(Y_{d_{i}})}-\frac{1}{30}g_{1}\beta{(g_{1})}-\frac{3}{2}g_{2}
\beta{(g_{2})}\right.\nonumber\\
&-&\left.\frac{8}{3}g_{3}\beta{(g_{3})}-2\left(\frac{x}{6}-\frac{1}{3}\right)^2g_{x}\beta(g_{x})\right],
\end{eqnarray}
\begin{eqnarray}
(m^2_{soft})_{\tilde{U}_i^c}^{\tilde{U}_i^c}
=\frac{M_{aux}^2}{16\pi^2}\left[2Y_{u_i}
\beta(Y_{u_i})-\frac{8}{15}g_1\beta(g_1)
-\frac{8}{3}g_3\beta(g_3)-2\left(-\frac{2x}{3}+\frac{1}{3}\right)^2g_{x}\beta(g_{x})\right],
\end{eqnarray}
\begin{eqnarray}
(m^{2}_{soft})_{\tilde{D}_{i}^{c}}^{\tilde{D}_{i}^{c}}=\frac{M_{aux}^{2}}{16\pi^2}\left[2Y_{d_{i}}
\beta{(Y_{d_{i}})}-\frac{2}{15}g_{1}\beta{(g_{1})}-\frac{8}{3}g_{3}\beta{(g_{3})}-2\left(\frac{x}{3}+\frac{1}{3}\right)^2g_{x}\beta(g_{x})\right].
\end{eqnarray}
\subsection{Heavy sneutrino masses}
The heavy right--handed sneutrinos ($\tilde{\nu}^c_i$) split into
scalar ($\tilde{\nu}_{is}^c$) and pseudoscalar
($\tilde{\nu}_{ip}^c$) components with masses given by
\begin{eqnarray}\label{A37}
m^2_{\tilde{\nu}_{is}^c}&=&\frac{M_{aux}^2}{(16\pi^2)}\left[4f_{\nu_{i}^c}\beta(f_{\nu_{i}^c})-2g_{x}\beta(g_{x})\right)]-2g_{x}^2(z^2-y^2)\nonumber\\
&+&2\mu^{\prime}f_{\nu_{i}^c}y+4f_{\nu_{i}^c}^2z^2+2f_{\nu_{i}^c}A_{\nu_i}z,\\
m^2_{\tilde{\nu}_{ip}^c}&=&\frac{M_{aux}^2}{(16\pi^2)}\left[4f_{\nu_{i}^c}\beta(f_{\nu_{i}^c})-2g_{x}\beta(g_{x})\right]-2g_{x}^2(z^2-y^2)\nonumber\\
&-&2\mu^{\prime}f_{\nu_{i}^c}y+4f_{\nu_{i}^c}^2z^2-2f_{\nu_{i}^c}A_{\nu_i}z.
\end{eqnarray}

As for the fourth heavy sneutrino, there is mixing between the
$\tilde{\nu}^c$ and the $\tilde{\bar{\nu}}^c$ fields. This leads
to two $2\times 2$ mass matrices, one for the scalars, and one for
the pseudoscalars. They are given by
\begin{eqnarray}\label{A380}
M^2_{\tilde{\nu}_s^c}&=&\pmatrix{m^2_{\tilde{\nu}_s^c}&2M_{\nu^c}\left(f_{\nu^c}z+hy+\frac{B_{\nu^c\bar{\nu}^c}}{2}\right)\cr
2M_{\nu^c}\left(f_{\nu^c}z+hy+\frac{B_{\nu^c\bar{\nu}^c}}{2}\right)&m^2_{\tilde{\bar{\nu}}_s^c}},\\
M^2_{\tilde{\nu}_p^c}&=&\pmatrix{m^2_{\tilde{\nu}_p^c}&2M_{\nu^c}\left(f_{\nu^c}z+hy+\frac{B_{\nu^c\bar{\nu}^c}}{2}\right)\cr
2M_{\nu^c}\left(f_{\nu^c}z+hy+\frac{B_{\nu^c\bar{\nu}^c}}{2}\right)&m^2_{\tilde{\bar{\nu}}_p^c}},
\end{eqnarray}
where
\begin{eqnarray}\label{A38}
m^2_{\tilde{\nu}_s^c}&=&\frac{M_{aux}^2}{(16\pi^2)}\left(4f_{\nu^c}\beta(f_{\nu^c})-2g_{x}\beta(g_{x})\right)-2g_{x}^2(z^2-y^2)\nonumber\\
&+&2\mu^{\prime}f_{\nu^c}y+4f_{\nu^c}^2z^2+2f_{\nu^c}A_{\nu^c}z+M_{\nu^c}^2,\\
m^2_{\tilde{\nu}_p^c}&=&\frac{M_{aux}^2}{(16\pi^2)}\left(4f_{\nu^c}\beta(f_{\nu^c})-2g_{x}\beta(g_{x})\right)-2g_{x}^2(z^2-y^2)\nonumber\\
&-&2\mu^{\prime}f_{\nu^c}y+4f_{\nu^c}^2z^2-2f_{\nu^c}A_{\nu^c}z+M_{\nu^c}^2,\\
m^2_{\tilde{\bar{\nu}}_s^c}&=&\frac{M_{aux}^2}{(16\pi^2)}\left(4h\beta(h)-2g_{x}\beta(g_{x})\right)+2g_{x}^2(z^2-y^2)\nonumber\\
&+&2\mu^{\prime}h z+4h^2y^2+2hA_{h}y+M_{\nu^c}^2,\\
m^2_{\tilde{\bar{\nu}}_p^c}&=&\frac{M_{aux}^2}{(16\pi^2)}\left(4h\beta(h)-2g_{x}\beta(g_{x})\right)+2g_{x}^2(z^2-y^2)\nonumber\\
&-&2\mu^{\prime}h z+4h^2y^2-2hA_{h}y+M_{\nu^c}^2,\\
B_{\nu^c\bar{\nu}^c}&=&-M_{aux}(\gamma_{\nu^c}+\gamma_{\bar{\nu}^c}).
\end{eqnarray}
Here $s$ ($p$) stands for scalar (pseudoscalar). The beta
functions, gamma functions and the $A$ terms are given in the
Appendix, Eqs. (125)--(\ref{A222}). We shall denote the mass
eigenstates of the scalars as
$\tilde{\omega}_{1s},\,\tilde{\omega}_{2s}$ with masses
$m^2_{\tilde{\omega}_{1s}},\,m^2_{\tilde{\omega}_{2s}}$, and the
pseudoscalars as $\tilde{\omega}_{1p},\,\tilde{\omega}_{2p}$ with
masses $m^2_{\tilde{\omega}_{1p}},\,m^2_{\tilde{\omega}_{2p}}$.

\section{Numerical Results for the Spectrum}
As inputs at $M_{Z}$ we choose the central values (in the
$\overline{MS}$ scheme ) \cite{Particle}
\begin{eqnarray}\label{A18}
\alpha_{3}(M_Z)=0.119,\,\,\,\,\sin^{2}{\theta_{W}}=0.23113,\,\,\,\,\alpha(M_Z)=\frac{1}{127.922}.
\end{eqnarray}
We keep the top quark mass fixed at its central value, $M_t=174.3$
GeV. We follow the procedure outlined in Ref. \cite{anoka} to
determine the parameter $\tan{\beta}$ and the lightest MSSM Higgs
boson mass $m_{h}$. The gauge couplings and the top quark Yukawa
coupling are evolved from the lower momentum scale to $Q=1$ TeV,
where the Higgs potential is minimized. We use the Standard Model
beta functions for this evolution. In determining the top quark
Yukawa coupling $Y_t(m_t)$, we use 2--loop QCD corrections to
convert the physical mass $M_t$ into the running mass $m_t(m_t)$.

For the lightest Higgs boson mass of MSSM we use the 2--loop
radiatively corrected expression for $
m^2_{h}=(m^2_{h})_{o}+\Delta m_{h}^2$, where $\Delta m_{h}^2$ is
given in Ref. \cite{Carena2}.

We present numerical results for two models: Model 1 with $x=1.3$,
and Model 2 with $x=1.6$. In Model 1, the left--handed sleptons
are heavier than the right--handed sleptons, while the reverse
holds for Model 2.

The value of $M_{aux}$ should be in the range $M_{aux}=40-100$ TeV
if the SUSY particles are to have masses in the range 100 GeV -- 2
TeV. In Table \ref{D3}, corresponding to Model 1, we choose
$M_{aux}=56.398$ TeV. In Table \ref{D5} (for Model 2) we choose
$M_{aux}= 59.987$ TeV. We have included the leading radiative
corrections \cite{Gunion} to $M_1,\,M_2$ and $M_3$ in our
numerical study. In Model 1 we find $M_1:M_2:M_3=3.0:1:7.1$. The
minimization conditions (Eqs. (\ref{A11})--(10)) fix
$\tan{\beta}=4.39$ in this model. The choice of $g_x=0.41$,
$f_{\nu_i^c}=f_{\nu^c}=0.28$, and $h=0.921$ are motivated by the
requirements of consistent symmetry breaking with $\langle
S_+\rangle\gtrsim \langle S_-\rangle\gg \upsilon_u,\,\upsilon_d$,
and the positivity of slepton masses. We find that the model
parameters are highly constrained. Only small deviations from the
choice in Table 2 are found to be consistent.

From Table \ref{D3} we see that the lightest Higgs boson of the
MSSM sector has mass of 121 GeV. The lightest SUSY particle is the
neutralino $\tilde{\chi}_1^0$, which is approximately a neutral
Wino. This is a candidate for cold dark matter \cite{cold}. Note
that $\tilde{\chi}_1^0$ is nearly mass degenerate with the lighter
chargino $\tilde{\chi}_1^\pm$ (which is approximately the charged
Wino). The mass splitting
$m_{\tilde{\chi}_1^0}-m_{\tilde{\chi}_1^\pm}=180$ MeV, where the
bulk (173 MeV) arises from finite electroweak radiative
corrections \cite{Pierce}, not shown in  Table \ref{D3}.

In the $U(1)_x$ sector, there is a relatively light neutral Higgs
boson $h^{\prime}$ with a mass of 60 GeV. This occurs since the
parameter $\tan{\psi}=\frac{z}{y}$ is close to 1 -- a requirement
for consistent symmetry breaking [see Eq. (\ref{A155})].
$h^{\prime }$ is an admixture of $S_+$ and $S_-$, and as such has
no direct couplings to the Standard Model fields. Its mass being
below 100 GeV is fully consistent with experimental constraints.
The phenomenology of such a weakly coupled light neutral Higgs
boson will be discussed in section 7.

The mass of the $Z^{\prime}$ gauge boson and the $Z-Z^{\prime}$
mixing angle are listed in Table 3 (for Model 1). In section 7 we
show that these values are compatible with known experimental
constraints.

Table 4 lists the eigenvectors of the neutralino mass matrix.
These will become relevant in discussing the decays of the
$Z^{\prime}$ gauge boson. Tables 5 and 6 give the eigenvectors of
the chargino and the CP--even Higgs bosons, which will also be
used in the study of $Z^{\prime}$ decays.

Tables \ref{D5}--11 are analogous to Tables \ref{D3}--6, except
that they now apply to Model 2 (with $x=1.6$). In this case,
$\tan{\beta}=5.83$ and $m_{h}=126$ GeV. Here the right--handed
sleptons are heavier than the left--handed sleptons. In fact, in
this Model, the LSP is the left--handed sneutrino. This can also
be a candidate for cold dark matter in the AMSB framework, as the
decay of the moduli fields and the gravitino will produce
$\tilde{\nu}_{Li}$ with an abundance of the right order
\cite{Pomarol,masiero}.

\newpage

\begin{center}
\begin{table}[h]
\begin{center}\small{
\begin{tabular}{|l|c|c|}\hline
\rule[1.5mm]{0mm}{0pt}Particles & Symbol& Mass (TeV)\\\hline
\rule[1.5mm]{0mm}{0pt}Neutralinos&$\{m_{\tilde{\chi}_{1}^{0}},\,\,m_{\tilde{\chi}_{2}^{0}},\,\,m_{\tilde{\chi}_{3}^{0}},\,\,m_{\tilde{\chi}_{4}^{0}}\}$&$\{0.175165,\,\,0.517,\,\,0.980,\,\,0.980\}$\\\hline
\rule[1.5mm]{0mm}{0pt}Neutralinos&$\{m_{\tilde{\chi}_{5}^{0}},\,\,m_{\tilde{\chi}_{6}^{0}},\,\,m_{\tilde{\chi}_{7}^{0}}\}$&$\{0.206,\,\,1.644,\,\,3.278\}$\\\hline
\rule[1.5mm]{0mm}{0pt}Charginos&$\{m_{\tilde{\chi}_{1}^{\pm}},\,\,m_{\tilde{\chi}_{2}^{\pm}}\}$&$\{0.175171,\,\,0.983\}$\\\hline
\rule[1.5mm]{0mm}{0pt}Gluino&$M_{3}$&$1.239$\\\hline
\rule[1.5mm]{0mm}{0pt}Neutral Higgs bosons
&$\{m_{h},\,\,m_{H},\,\,m_{A}\}$&$\{0.121,\,\,0.793,\,\,0.792\}$\\\hline
\rule[1.5mm]{0mm}{0pt}Neutral Higgs bosons
&$\{m_{h^\prime},\,\,m_{H^\prime},\,\,m_{A^\prime}\}$&$\{0.060,\,\,2.394,\,\,0.241\}$\\\hline
\rule[1.5mm]{0mm}{0pt}Charged Higgs bosons
&$m_{H^{\pm}}$&$0.796$\\\hline \rule[1.5mm]{0mm}{0pt}R.H sleptons
&$\{m_{\tilde{e}_{R}},\,\,m_{\tilde{\mu}_{R}},\,\,m_{\tilde{\tau}_{1}}\}$&$\{0.215,\,\,0.215,\,\,0.205\}$\\\hline
\rule[1.5mm]{0mm}{0pt}L.H sleptons
&$\{m_{\tilde{e}_{L}},\,\,m_{\tilde{\mu}_{L}},\,\,m_{\tilde{\tau}_{2}}\}$&$\{0.249,\,\,0.249,\,\,0.257\}$\\\hline
\rule[1.5mm]{0mm}{0pt}Sneutrinos&$\{m_{\tilde{\nu}_{e}},\,\,m_{\tilde{\nu}_{\mu}},\,\,m_{\tilde{\nu}_{\tau}}\}$&$\{0.220,\,\,0.220,\,\,0.220\}$\\\hline
\rule[1.5mm]{0mm}{0pt}R.H down squarks
&$\{m_{\tilde{d}_{R}},\,\,m_{\tilde{s}_{R}},\,\,m_{\tilde{b}_{1}}\}$&$\{1.284,\,\,1.284,\,\,1.284\}$\\\hline
\rule[1.5mm]{0mm}{0pt}L.H down squarks
&$\{m_{\tilde{d}_{L}},\,\,m_{\tilde{s}_{L}},\,\,m_{\tilde{b}_{2}}\}$&$\{1.186,\,\,1.186,\,\,1.028\}$\\\hline
\rule[1.5mm]{0mm}{0pt}R.H up squarks
&$\{m_{\tilde{u}_{R}},\,\,m_{\tilde{c}_{R}},\,\,m_{\tilde{t}_{1}}\}$&$\{1.098,\,\,1.098,\,\,0.644\}$\\\hline
\rule[1.5mm]{0mm}{0pt}L.H up squarks
&$\{m_{\tilde{u}_{L}},\,\,m_{\tilde{c}_{L}},\,\,m_{\tilde{t}_{2}}\}$&$\{1.184,\,\,1.184,\,\,1.099\}$\\
\hline
 \rule[1.5mm]{0mm}{0pt}R.H scalar neutrinos
&$\{m_{\tilde{\nu}_{s_i}^c}\}$($i=1-3$)&$0.605$\\\hline
\rule[1.5mm]{0mm}{0pt}R.H pseudoscalar neutrinos
&$\{m_{\tilde{\nu}_{p_i}^c}\}$($i=1-3$)&$0.413$\\\hline

\rule[1.5mm]{0mm}{0pt}Heavy scalar neutrino
($\tilde{\nu}^c,\,\,\tilde{\bar{\nu}}^c$)
&$\{m_{{\tilde{\omega}}_{1s}},\,m_{{\tilde{\omega}}_{2s}}\}$&$\{1.142,\,\,\,3.644\}$\\\hline

\rule[1.5mm]{0mm}{0pt}Heavy pseudoscalar neutrino
($\tilde{\nu}^c,\,\,\tilde{\bar{\nu}}^c$)
&$\{m_{{\tilde{\omega}}_{ps}},\,m_{{\tilde{\omega}}_{2p}}\}$&$\{0.595,\,\,\,1.439\}$\\\hline

\rule[1.5mm]{0mm}{0pt}R.H  neutrinos
&$\{m_{\nu^c_i}\}$&$0.455$\\\hline

\rule[1.5mm]{0mm}{0pt}Heavy neutrinos ($\nu^c,\,\,\bar{\nu}^c$)
&$\{m_{\omega_1},\,\,m_{\omega_2}\}$&$\{0.933,\,\,1.635\}$\\\hline

\end{tabular}
\caption{\footnotesize \small{Sparticle masses in Model 1
($x=1.3$) for the choice $M_{aux}=56.398$ TeV,
$\tan{\psi}=-1.295$, $u=2.054$ TeV, $f_{\nu_{i}^c}=0.28$,
$f_{\nu^c}=0.28$, $h=0.921$, $g_{x}=0.41$, $M_{\nu^c}=1$ TeV and
$M_t=174.3$ GeV. This corresponds to $\tan{\beta}=4.39$, $\mu
=-0.977$ TeV, $\mu^{\prime} =0.214$ TeV, $y_{b}=0.03$.}}
  \label{D3}}
  \end{center}
\end{table}
\end{center}

\begin{table}[!h]
\begin{center}\small{
\begin{tabular}{|l|c|c|}\hline
\rule[1.5mm]{0mm}{0pt}$Z^{\prime}$ boson
mass&$M_{Z^{\prime}}$&$2.383$ TeV\\\hline
\rule[1.5mm]{0mm}{0pt}$Z-Z^{\prime}$ mixing angle
&$\xi$&$0.001$\\\hline
\end{tabular}
\caption{\footnotesize \small{$Z^{\prime}$ mass and $Z-Z^{\prime}$
mixing angle in Model 1 for the same set of input parameters as in
Table \ref{D3}.}}
  \label{D31}}
  \end{center}
\end{table}

\newpage
~\vspace{.3in}
\begin{center}
\begin{table}[!h]
\begin{center}\small{
\begin{tabular}{|c|c|c|c|c|c|c|c|}\hline
\rule[1.5mm]{0mm}{0pt}Fields& $\tilde{\chi}^0_1$&
$\tilde{\chi}^0_2$& $\tilde{\chi}^0_3$& $\tilde{\chi}^0_4$&
$\tilde{\chi}^0_5$& $\tilde{\chi}^0_6$& $\tilde{\chi}^0_7$\\\hline
\rule[1.5mm]{0mm}{0pt}$\tilde{B}$&-0.003&0.998&0.051&0.025&0.000&-0.001&0.000\\\hline
\rule[1.5mm]{0mm}{0pt}$\tilde{W}_3^0$&-0.997&0.001&-0.052&-0.058&0.000&0.002&0.000\\\hline
\rule[1.5mm]{0mm}{0pt}$\tilde{H}^0_d$&0.078&0.054&-0.703&-0.704&-0.002&0.030&0.001\\\hline
\rule[1.5mm]{0mm}{0pt}$\tilde{H}^0_u$&-0.004&0.019&-0.707&0.706&0.001&-0.042&0.016\\\hline
\rule[1.5mm]{0mm}{0pt}$\tilde{B}^{\prime}$&0.000&0.000&-0.004&-0.023&-0.026&-0.612&-0.790\\\hline
\rule[1.5mm]{0mm}{0pt}$\tilde{S}_+$&0.000&0.000&-0.011&0.039&-0.597&0.642&-0.479\\\hline
\rule[1.5mm]{0mm}{0pt}$\tilde{S}_-$&0.000&0.000&-0.009&0.026&0.802&0.458&-0.382\\\hline
\end{tabular}
\caption{\footnotesize \small{Eigenvectors of the neutralino mass
matrix in Model 1. The unitary matrix $O$ in Eq. (\ref{A311}) is
the transpose of this array. }}
  \label{D7}}
  \end{center}
\end{table}
\end{center}

\begin{center}
\begin{table}[!h]
\begin{center}\small{
\begin{tabular}{|c|c|c|c|c|c|c|c|}\hline
\rule[1.5mm]{0mm}{0pt} $U_{11}$& $U_{12}$& $U_{21}$& $U_{22}$&
$V_{11}$& $V_{12}$& $V_{21}$& $V_{22}$\\\hline
\rule[1.5mm]{0mm}{0pt}0.994&0.110&-0.110&0.994&1.000&0.006&-0.006&1.000\\\hline

\end{tabular}
\caption{\footnotesize \small{Eigenvectors of the chargino mass
matrix in Model 1, where $U$, $V$ are the unitary matrices that
diagonalize the chargino mass matrix
($V^{\ast}M^{(c)}U^{-1}=M^{(c)}_{diag}$). }}
  \label{D17}}
  \end{center}
\end{table}
\end{center}

\begin{center}
\begin{table}[!h]
\begin{center}\small{
\begin{tabular}{|c|c|c|c|c|}\hline
\rule[1.5mm]{0mm}{0pt}Fields& $h$& $h^{\prime}$& $H$&
$H^{\prime}$\\\hline
\rule[1.5mm]{0mm}{0pt}$H^0_d$&0.226&-0.025&0.974&-0.007\\\hline
\rule[1.5mm]{0mm}{0pt}$H^0_u$&0.967&-0.110&-0.227&0.027\\\hline
\rule[1.5mm]{0mm}{0pt}$S_+$&-0.050&-0.612&-0.010&-0.790\\\hline
\rule[1.5mm]{0mm}{0pt}$S_-$&0.104&0.783&-0.008&-0.613\\\hline

\end{tabular}
\caption{\footnotesize \small{Eigenvectors of the CP--even Higgs
boson mass matrix in Model 1. This array corresponds to $X$ used
in Eqs. (82) -- (84) and Eq. (109) of the text.}}
  \label{D8}}
  \end{center}
\end{table}
\end{center}

\newpage
\begin{center}
\begin{table}[!h]
\begin{center}\small{
\begin{tabular}{|l|c|c|}\hline
\rule[1.5mm]{0mm}{0pt}Particles & Symbol& Mass (TeV)\\\hline
\rule[1.5mm]{0mm}{0pt}Neutralinos&$\{m_{\tilde{\chi}_{1}^{0}},\,\,m_{\tilde{\chi}_{2}^{0}},\,\,m_{\tilde{\chi}_{3}^{0}},\,\,m_{\tilde{\chi}_{4}^{0}}\}$&$\{0.185.851,\,\,0.550,\,\,1.049,\,\,1.050\}$\\\hline
\rule[1.5mm]{0mm}{0pt}Neutralinos&$\{m_{\tilde{\chi}_{5}^{0}},\,\,m_{\tilde{\chi}_{6}^{0}},\,\,m_{\tilde{\chi}_{7}^{0}}\}$&$\{0.498,\,\,2.840,\,\,4.539\}$\\\hline
\rule[1.5mm]{0mm}{0pt}Charginos&$\{m_{\tilde{\chi}_{1}^{\pm}},\,\,m_{\tilde{\chi}_{2}^{\pm}}\}$&$\{0.185855,\,\,1.051\}$\\\hline
\rule[1.5mm]{0mm}{0pt}Gluino&$M_{3}$&$1.298$\\\hline
\rule[1.5mm]{0mm}{0pt}Neutral Higgs bosons
&$\{m_{h},\,\,m_{H},\,\,m_{A}\}$&$\{0.126,\,\,0.625,\,\,0.625\}$\\\hline
\rule[1.5mm]{0mm}{0pt}Neutral Higgs bosons
&$\{m_{h^\prime},\,\,m_{H^\prime},\,\,m_{A^\prime}\}$&$\{0.023,\,\,3.436,\,\,0.125\}$\\\hline
\rule[1.5mm]{0mm}{0pt}Charged Higgs bosons
&$m_{H^{\pm}}$&$0.630$\\\hline \rule[1.5mm]{0mm}{0pt}R.H sleptons
&$\{m_{\tilde{e}_{R}},\,\,m_{\tilde{\mu}_{R}},\,\,m_{\tilde{\tau}_{1}}\}$&$\{0.383,\,\,0.383,\,\,0.385\}$\\\hline
\rule[1.5mm]{0mm}{0pt}L.H sleptons
&$\{m_{\tilde{e}_{L}},\,\,m_{\tilde{\mu}_{L}},\,\,m_{\tilde{\tau}_{2}}\}$&$\{0.213,\,\,0.213,\,\,0.210\}$\\\hline
\rule[1.5mm]{0mm}{0pt}Sneutrinos&$\{m_{\tilde{\nu}_{e}},\,\,m_{\tilde{\nu}_{\mu}},\,\,m_{\tilde{\nu}_{\tau}}\}$&$\{0.174,\,\,0.174,\,\,0.174\}$\\\hline
\rule[1.5mm]{0mm}{0pt}R.H down squarks
&$\{m_{\tilde{d}_{R}},\,\,m_{\tilde{s}_{R}},\,\,m_{\tilde{b}_{1}}\}$&$\{1.370,\,\,1.370,\,\,1.369\}$\\\hline
\rule[1.5mm]{0mm}{0pt}L.H down squarks
&$\{m_{\tilde{d}_{L}},\,\,m_{\tilde{s}_{L}},\,\,m_{\tilde{b}_{2}}\}$&$\{1.267,\,\,1.267,\,\,1.087\}$\\\hline
\rule[1.5mm]{0mm}{0pt}R.H up squarks
&$\{m_{\tilde{u}_{R}},\,\,m_{\tilde{c}_{R}},\,\,m_{\tilde{t}_{1}}\}$&$\{1.031,\,\,1.031,\,\,0.406\}$\\\hline
\rule[1.5mm]{0mm}{0pt}L.H up squarks
&$\{m_{\tilde{u}_{L}},\,\,m_{\tilde{c}_{L}},\,\,m_{\tilde{t}_{2}}\}$&$\{1.264,\,\,1.264,\,\,1.1141\}$\\
\hline
 \rule[1.5mm]{0mm}{0pt}R.H scalar neutrinos
&$\{m_{\tilde{\nu}_{s_i}^c}\}$($i=1-3$)&$1.583$\\\hline

\rule[1.5mm]{0mm}{0pt}R.H pseudoscalar neutrinos
&$\{m_{\tilde{\nu}_{p_i}^c}\}$($i=1-3$)&$1.129$\\\hline

\rule[1.5mm]{0mm}{0pt}Heavy scalar neutrino
($\tilde{\nu}^c,\,\,\tilde{\bar{\nu}}^c$)
&$\{m_{{\tilde{\omega}}_{1s}},\,m_{{\tilde{\omega}}_{2s}}\}$&$\{1.852,\,\,\,4.700\}$\\\hline

\rule[1.5mm]{0mm}{0pt}Heavy pseudoscalar neutrino
($\tilde{\nu}^c,\,\,\tilde{\bar{\nu}}^c$)
&$\{m_{{\tilde{\omega}}_{ps}},\,m_{{\tilde{\omega}}_{2p}}\}$&$\{1.398,\,\,\,2.586\}$\\\hline

\rule[1.5mm]{0mm}{0pt}R.H  neutrinos
&$\{m_{\nu^c_i}\}$&$0.829$\\\hline

\rule[1.5mm]{0mm}{0pt}Heavy neutrinos ($\nu^c,\,\,\bar{\nu}^c$)
&$\{m_{\omega_1},\,\,m_{\omega_2}\}$&$\{1.174,\,\,2.070\}$\\\hline
\end{tabular}
\caption{\footnotesize \small{Sparticle masses in Model 2
($x=1.6$) for the choice $M_{aux}=59.987$ TeV,
$\tan{\psi}=-1.202$,
 $u=2.697$ TeV,
$f_{\nu_{i}^c}=0.4$, $f_{\nu^c}=0.4$,
 $h=1.0$, $g_{x}=0.45$, $M_1^{\prime}=2.197$ TeV, $M_{\nu^c}=1$ TeV and $M_t=174.3$ GeV. This corresponds to $\tan{\beta}=5.83$, $\mu
=-1.046$ TeV, $\mu^{\prime} =-0.505$ TeV, $y_{b}=0.06$.}}
  \label{D5}}
  \end{center}
\end{table}
\end{center}

\begin{table}[!h]
\begin{center}\small{
\begin{tabular}{|l|c|c|}\hline

 \rule[1.5mm]{0mm}{0pt}$Z^{\prime}$ boson mass&$M_{Z^{\prime}}$&$3.433$ TeV\\\hline
\rule[1.5mm]{0mm}{0pt}$Z-Z^{\prime}$ mixing angle
&$\xi$&$0.00068$\\\hline
\end{tabular}
\caption{\footnotesize \small{$Z^{\prime}$ mass and $Z-Z^{\prime}$
mixing angle in Model 2 for the same set of input parameters as in
Table \ref{D5}.}}
  \label{D51}}
  \end{center}
\end{table}

\newpage
~\vspace{.3in}
\begin{center}
\begin{table}[!h]
\begin{center}\small{
\begin{tabular}{|c|c|c|c|c|c|c|c|}\hline
\rule[1.5mm]{0mm}{0pt}Fields& $\tilde{\chi}^0_1$&
$\tilde{\chi}^0_2$& $\tilde{\chi}^0_3$& $\tilde{\chi}^0_4$&
$\tilde{\chi}^0_5$& $\tilde{\chi}^0_6$& $\tilde{\chi}^0_7$\\\hline
\rule[1.5mm]{0mm}{0pt}$\tilde{B}$&-0.001&0.998&-0.052&0.023&0.000&0.000&0.000\\\hline
\rule[1.5mm]{0mm}{0pt}$\tilde{W}_3^0$&-0.997&0.002&0.053&-0.052&0.000&-0.001&0.000\\\hline
\rule[1.5mm]{0mm}{0pt}$\tilde{H}^0_d$&-0.074&-0.052&-0.703&0.705&-0.002&0.011&0.001\\\hline
\rule[1.5mm]{0mm}{0pt}$\tilde{H}^0_u$&0.000&-0.020&-0.707&-0.707&-0.001&-0.021&0.016\\\hline
\rule[1.5mm]{0mm}{0pt}$\tilde{B}^{\prime}$&0.000&0.000&0.006&-0.004&0.023&0.0563&0.826\\\hline
\rule[1.5mm]{0mm}{0pt}$\tilde{S}_+$&0.000&0.000&0.011&0.018&-0.648&-0.620&0.441\\\hline
\rule[1.5mm]{0mm}{0pt}$\tilde{S}_-$&0.000&0.000&0.007&0.017&0.761&-0.546&0.350\\\hline
\end{tabular}
\caption{\footnotesize \small{Eigenvectors of the neutralino mass
matrix in Model 2. The unitary matrix $O$ in Eq. (\ref{A311}) is
the transpose of this array.}}
  \label{D9}}
  \end{center}
\end{table}
\end{center}

\begin{center}
\begin{table}[!h]
\begin{center}\small{
\begin{tabular}{|c|c|c|c|c|c|c|c|}\hline
\rule[1.5mm]{0mm}{0pt} $U_{11}$& $U_{12}$& $U_{21}$& $U_{22}$&
$V_{11}$& $V_{12}$& $V_{21}$& $V_{22}$\\\hline
\rule[1.5mm]{0mm}{0pt}0.994&0.105&-0.105&0.994&1.000&0.000&-0.000&1.000\\\hline

\end{tabular}
\caption{\footnotesize \small{Eigenvectors of the chargino mass
matrix in Model 2, where $U$, $V$ are the unitary matrices that
diagonalize the chargino mass matrix
($V^{\ast}M^{(c)}U^{-1}=M^{(c)}_{diag}$). }}
  \label{D18}}
  \end{center}
\end{table}
\end{center}

\begin{center}
\begin{table}[!h]
\begin{center}\small{
\begin{tabular}{|c|c|c|c|c|}\hline
\rule[1.5mm]{0mm}{0pt} Fields& $h$& $h^{\prime}$& $H$&
$H^{\prime}$\\\hline
\rule[1.5mm]{0mm}{0pt}$H^0_d$&0.176&0.002&0.984&0.005\\\hline
\rule[1.5mm]{0mm}{0pt}$H^0_u$&0.984&0.010&-0.176&-0.025\\\hline
\rule[1.5mm]{0mm}{0pt}$S_+$&-0.012&-0.640&0.007&-0.768\\\hline
\rule[1.5mm]{0mm}{0pt}$S_-$&-0.023&0.768&0.006&-0.640\\\hline

\end{tabular}
\caption{\footnotesize \small{Eigenvectors of the CP--even Higgs
boson mass matrix in Model 2. This array corresponds to $X$ used
in Eqs. (82) -- (84) and Eq. (109) of the text.}}
  \label{D10}}
  \end{center}
\end{table}
\end{center}

\newpage

\section{$Z^{\prime} $ Decay Modes and Branching Ratios}
The $Z^{\prime}$ gauge boson of our model has substantial coupling
to the quarks. With its mass in the range 2--4 TeV, it will be
produced copiously at the LHC via the process $pp\rightarrow
Z^{\prime}$. The reach of LHC is about 5 TeV for a $Z^{\prime}$
with generic quark and lepton couplings \cite{djouadi}. Our model
will then be directly tested at the LHC. Once produced, the
$Z^{\prime}$ will decay into various channels. It is important to
identify the dominant decay modes of the $Z^{\prime}$ and
calculate the corresponding branching ratios. This is what we do
in this section. We will see that our $Z^{\prime}$ is almost
leptophobic, with $Br(Z^{\prime}\rightarrow e^+e^-)=(1-1.6)\%$.
Direct limits on such a $Z^{\prime}$ are rather weak, however, the
$Z-Z^{\prime}$ mixing which occurs in our models at the level of
0.001 does provide useful constraints.

We now turn to the dominant 2--body decays of $Z^{\prime}$. In
this analysis we can safely ignore the small $Z-Z^{\prime}$ mixing
for the most part.

The Lagrangian for $Z^{\prime}$ coupling to the Standard Model
fermions can be written as
\begin{eqnarray}\label{A300}
\mathcal{L}=g_x\bar{f}\gamma^\mu(v_f-a_f\gamma_5)fZ_{\mu}^{\prime}.
\end{eqnarray}
The $Z^{\prime}$ decay rate into a fermion--antifermion pair is
then
\begin{eqnarray}\label{A400}
\Gamma(Z^{\prime}\rightarrow
\bar{f}f)=C_{f}\frac{g_{x}^2}{12\pi}M_{Z^{\prime}}\left[v^2_{f}\left(1+2\frac{m_f^2}{M_{Z^{\prime}}^2}\right)+a_f^2\left(1-4\frac{m_f^2}{M_{Z^{\prime}}^2}\right)\right]\sqrt{1-4\frac{m_f^2}{M_{Z^{\prime}}^2}}.
\end{eqnarray}
Here $C_f=3\,\,(1)$ for quarks (leptons), $M_{Z^{\prime}}$ is the
$Z^{\prime}$ mass and $g_{x}$ is the $U(1)_{x}$ gauge coupling.
The vector and the axial--vector couplings ($v_f,\,\,a_f$) are
related to the $U(1)_x$ charges of the fermions as
\begin{eqnarray}\label{A401}
v_f&=&\frac{1}{2}\left(Q(f_L)+Q(f_R)\right),\\
a_f&=&\frac{1}{2}\left(Q(f_L)-Q(f_R)\right).
\end{eqnarray}
Here $Q$ is the $U(1)_x$ charge of $f_L$ (listed in Table 1 ) and
$Q(f_R)=-Q(f_L^c)$.

The decay width for $Z^{\prime}\rightarrow \bar{\nu}_{Li}\nu_{Li}$
and $Z^{\prime}\rightarrow \bar{\nu}_{i}^c\nu_{i}^c$ are:
\begin{eqnarray}\label{A403}
\Gamma(Z^{\prime}\rightarrow
\bar{\nu}_{Li}\nu_{Li})&=&\frac{g_{x}^2}{24\pi}Q_{\nu_{Li}}^2M_{Z^{\prime}},\\
\Gamma(Z^{\prime}\rightarrow
\bar{\nu}^c_i\nu_i^c)&=&\frac{g_{x}^2}{24\pi}Q_{\nu_i^c}^2M_{Z^{\prime}}\left(1-4\frac{m_{\nu_i^c}^2}{M_{Z^{\prime}}^2}\right)^{\frac{3}{2}}.
\end{eqnarray}

 There is mixing between the heavy vector--like $\nu^c$ and the
$\bar{\nu}^c$ [Cf: Eq. (\ref{A382})], with the mass eigenstates
($\omega_1,\,\omega_2$) given by
\begin{eqnarray}\label{A4033}
\pmatrix{\nu^c\cr
\bar{\nu}^c}=\pmatrix{\cos{\theta_{\nu^c}}&\sin{\theta_{\nu^c}}\cr
-\sin{\theta_{\nu^c}}&\cos{\theta_{\nu^c}}}\pmatrix{\omega_1\cr
\omega_2}.
\end{eqnarray}
Since $Q_{\bar{\nu}^c}=-Q_{\nu^c}$, the Lagrangian for the
$Z^{\prime}$ coupling to these neutrino is given by
\begin{eqnarray}\label{A301}
\mathcal{L}&=&\frac{g_x}{2}Q_{\nu^c}(
\cos{2\theta_{\nu^c}}\bar{\omega}_1\gamma^\mu\gamma_5\omega_1-
\cos{2\theta_{\nu^c}}\bar{\omega}_2\gamma^\mu\gamma_5\omega_2-
\sin{2\theta_{\nu^c}}\bar{\omega}_1\gamma^\mu\gamma_5\omega_2\nonumber\\&-&
\sin{2\theta_{\nu^c}}\bar{\omega}_2\gamma^\mu\gamma_5\omega_1)Z_{\mu}^{\prime}.
\end{eqnarray}
This leads to the decay rates
\begin{eqnarray}\label{A4034}
\Gamma(Z^{\prime}\rightarrow
{\omega}_1\omega_1)&=&\frac{g_{x}^2}{24\pi}M_{Z^{\prime}}Q_{\nu^c}^2\cos^2{2\theta_{\nu^c}}\left(1-4\frac{m_{\omega_1}^2}{M_{Z^{\prime}}^2}\right)^{\frac{3}{2}},\\
\Gamma(Z^{\prime}\rightarrow
{\omega}_2\omega_2)&=&\frac{g_{x}^2}{24\pi}M_{Z^{\prime}}Q_{\nu^c}^2\cos^2{2\theta_{\nu^c}}\left(1-4\frac{m_{\omega_2}^2}{M_{Z^{\prime}}^2}\right)^{\frac{3}{2}},\\
\Gamma(Z^{\prime}\rightarrow
{\omega}_1\omega_2)&=&\frac{g_{x}^2}{12\pi}M_{Z^{\prime}}Q_{\nu^c}^2\sin^2{2\theta_{\nu^c}}\left[1-\frac{(m_{\omega_1}^2+m_{\omega_2}^2)}{2M_{Z^{\prime}}^2}
-\frac{(m_{\omega_1}^2-m_{\omega_2}^2)^2}{2M_{Z^{\prime}}^4}-3\frac{m_{\omega_1}m_{\omega_2}}{M_{Z^{\prime}}^2}\right]\nonumber\\
&\times&\sqrt{\left(1-\frac{(m_{\omega_1}+m_{\omega_2})^2}{M_{Z^{\prime}}^2}\right)\left(1-\frac{(m_{\omega_1}-m_{\omega_2})^2}{M_{Z^{\prime}}^2}\right)}.
\end{eqnarray}
Here $m_{\omega_1}\,\,(m_{\omega_2})$ are the masses of the
physical Majorana fermions.

The $Z^{\prime}$ interaction with the sfermions is described by
the Lagrangian
\begin{eqnarray}\label{A303}
\mathcal{L}=ig_x(v_f\pm
a_f)\tilde{f}_{L,R}^{*}\stackrel{\leftrightarrow}{\partial}_\mu
\tilde{f}_{L,R}Z^{\prime \mu}.
\end{eqnarray}
The rate for the decay $Z^{\prime}$ to sfermions is given by
\begin{eqnarray}\label{A402}
\Gamma(Z^{\prime}\rightarrow
\tilde{f}_{L,R}^{*}\tilde{f}_{L,R})=C_{f}\frac{g_{x}^2}{48\pi}M_{Z^{\prime}}(v_{f}\pm
a_f)^2\left(1-4\frac{m_{\tilde{f}_{L,R}}^2}{M_{Z^{\prime}}^2}\right)^{\frac{3}{2}},
\end{eqnarray}
where the $+(-)$ sign is for the left (right)--handed sfermions
and $m_{\tilde{f}_{L,R}}$ is the left (right)--handed sfermion
mass. $v_f$ and $a_f$ are as given in Eqs. (\ref{A401})--(59).

In the top squark sector, there is non--negligible mixing between
the left and the right--handed sfermions. This leads to the
following modification of the Lagrangian:
\begin{eqnarray}\label{A325}
 \mathcal{L}=ig_x\left((v_f\pm a_f\cos{2\theta_{\tilde{f}}})\tilde{f}^{*}_{1,2}\stackrel{\leftrightarrow}{\partial}_\mu \tilde{f}_{1,2}
 -a_f\sin{2\theta_{\tilde{f}}}(\tilde{f}^{*}_1\stackrel{\leftrightarrow}{\partial}_\mu \tilde{f}_2+\tilde{f}^{*}_2\stackrel{\leftrightarrow}{\partial}_\mu \tilde{f}_1)\right){{Z^\prime}^\mu},
\end{eqnarray}
where $\theta_{\tilde{f}}$ is the left--right sfermion mixing
angle. The decay rate is given by
\begin{eqnarray}\label{A4021}
\Gamma(Z^{\prime}\rightarrow
\tilde{f}_{1,2}^{*}\tilde{f}_{1,2})&=&C_{f}\frac{g_{x}^2}{48\pi}M_{Z^{\prime}}(v_{f}\pm
a_f\cos{2\theta_{\tilde{f}}})^2\left(1-4\frac{m_{\tilde{f}_{1,2}}^2}{M_{Z^{\prime}}^2}\right)^{\frac{3}{2}},\\
\Gamma(Z^{\prime}\rightarrow
\tilde{f}_{1}^{*}\tilde{f}_{2})&=&C_{f}\frac{g_{x}^2}{48\pi}M_{Z^{\prime}}(a_f\sin{2\theta_{\tilde{f}}})^2
\left[1+2\frac{(m_1^2+m_2^2)}{M^2_{Z^{\prime}}}+\frac{(m_1^2-m_2^2)^2}{M^4_{Z^{\prime}}})\right]^{\frac{3}{2}}.
\end{eqnarray}

The $\tilde\nu^c$ and  $\tilde{\bar{\nu}}^c$ splits into two
scalar and two pseudoscalar which mix (see Eqs.
(\ref{A380})--(49)). The mass eigenstate  $\tilde{\omega}_{is}$
and $\tilde{\omega_{ip}}$ are given as
\begin{eqnarray}\label{A500}
\pmatrix{\tilde{\nu}_s^c\cr
\tilde{\bar{\nu}}_s^c}&=&\pmatrix{\cos{\theta_{\omega
s}}&\sin{\theta_{\omega s}}\cr -\sin{\theta_{\omega
s}}&\cos{\theta_{\omega s}}}\pmatrix{\tilde{\omega}_{1s}\cr
\tilde{\omega}_{2s}},\\\nn\\
\pmatrix{\tilde{\nu}_p^c\cr
\tilde{\bar{\nu}}_p^c}&=&\pmatrix{\cos{\theta_{\omega
p}}&\sin{\theta_{\omega p}}\cr -\sin{\theta_{\omega
p}}&\cos{\theta_{\omega p}}}\pmatrix{\tilde{\omega}_{1p}\cr
\tilde{\omega}_{2p}}.
\end{eqnarray}
The Lagrangian for the $Z^{\prime}$ coupling to the
scalar--pseudoscalar pair is given by:
\begin{eqnarray}\label{A501}
\mathcal{L}&=&g_x\left[(Q_{\nu^c}\cos{\theta_{\omega
s}}\cos{\theta_{\omega p}}+Q_{\bar{\nu}^c}\sin{\theta_{\omega
s}}\sin{\theta_{\omega
p}})\tilde{\omega}_{1s}\stackrel{\leftrightarrow}{\partial}_\mu
\tilde{\omega}_{1p}
\right.\nonumber\\
&&\left.+(Q_{\nu^c}\sin{\theta_{\omega s}}\sin{\theta_{\omega p}}+Q_{\bar{\nu}^c}\cos{\theta_{\omega s}}\cos{\theta_{\omega p}})\tilde{\omega}_{2s}\stackrel{\leftrightarrow}{\partial}_\mu \tilde{\omega}_{2p}\right.\nonumber\\
&&\left.+(Q_{\nu^c}\cos{\theta_{\omega s}}\sin{\theta_{\omega p}}-Q_{\bar{\nu}^c}\sin{\theta_{\omega s}}\cos{\theta_{\omega p}})\tilde{\omega}_{1s}\stackrel{\leftrightarrow}{\partial}_\mu \tilde{\omega}_{2p}\right.\nonumber\\
&&\left.+ (Q_{\nu^c}\sin{\theta_{\omega s}}\cos{\theta_{\omega
p}}-Q_{\bar{\nu}^c}\cos{\theta_{\omega s}}\sin{\theta_{\omega
p}})\tilde{\omega}_{2s}\stackrel{\leftrightarrow}{\partial}_\mu
\tilde{\omega}_{1p}\right]{{Z^\prime}^\mu}.
\end{eqnarray}
This leads to the decay rate
\begin{eqnarray}\label{A502}
\Gamma(Z^{\prime}\rightarrow
\tilde{\omega}_{is}\tilde{\omega}_{jp})=\frac{g_{x}^2}{48\pi}Q_{ij}^2\left[1-2\frac{(m_{\omega_{is}}^2+m_{\omega_{jp}}^2)}{M_{Z^{\prime}}^2}+\frac{(m_{\omega_{is}}^2-m_{\omega_{jp}}^2)^2}{M_{Z^{\prime}}^4}\right]^{\frac{3}{2}},
\end{eqnarray}
where $Q_{ij}$ is identified with the appropriate coupling to
$\tilde{\omega}_{is}\tilde{\omega}_{jp}$ term in the Lagrangian of
Eq. (\ref{A501}).

The supersymetric partners of $\nu^c_{i}$ split into a scalar
($\tilde{\nu}_{i s}^c$) and a pseudoscalar ($\tilde{\nu}_{i
p}^c$). The decay of $Z^{\prime}$ to these fields is similar to
those analyzed in Eq. (\ref{A502}):
\begin{eqnarray}\label{A5021}
\Gamma(Z^{\prime}\rightarrow
\tilde{\nu}^c_{is}\tilde{\nu}^c_{ip})=\frac{g_{x}^2}{48\pi}Q_{\nu_i^c}^2\left[1-2\frac{(m_{\tilde{\nu}^c_{is}}^2+m_{\tilde{\nu}^c_{ip}}^2)}{M_{Z^{\prime}}^2}+\frac{(m_{\tilde{\nu}^c_{is}}^2-m_{\tilde{\nu}^c_{ip}}^2)^2}{M_{Z^{\prime}}^4}\right]^{\frac{3}{2}},
\end{eqnarray}
where $m_{\tilde{\nu}^c_{is}}$ and $m_{\tilde{\nu}^c_{ip}}$ are
the masses of the scalar and the pseudoscalar.

 The Lagrangian for the $Z^{\prime}$
coupling to the charged Higgs bosons is given by
\begin{eqnarray}\label{A304}
\mathcal{L}&=&ig_x(Q_{H_d}\sin^2{\beta}-Q_{H_u}\cos^2{\beta})H^+\stackrel{\leftrightarrow}{\partial}_\mu H^-{{Z^\prime}^\mu}\nonumber\\
&+&g_x(Q_{H_d}+Q_{H_u})\sin{\beta}\cos{\beta}M_W(W^+_{\mu
}H^-+W^{-}_\mu H^+){{Z^\prime}^\mu},
\end{eqnarray}
where $Q_{H_d}$ $(Q_{H_u})$ is the $U(1)_{x}$ charge of $H_{d}$
($H_u$) field. The decay rates of $Z^{\prime}$ to $H^+H^-$ and
$W^{\pm}H^{\mp}$ are given by
\begin{eqnarray}\label{A404}
\Gamma(Z^{\prime}\rightarrow
H^+H^-)&=&\frac{g_{x}^2}{48\pi}M_{Z^{\prime}}(Q_{H_d}\sin^2{\beta}-Q_{H_u}\cos^2{\beta})^2\left(1-4\frac{m_{H^{\pm}}^2}{M_{Z^{\prime}}^2}\right)^{\frac{3}{2}},\\
\Gamma(Z^{\prime}\rightarrow
W^{\pm}H^{\mp})&=&\frac{g_{x}^2}{192\pi}M_{Z^{\prime}}(Q_{H_d}+Q_{H_u})^2\left[1+2\frac{(5M_W^2-m^2_{H^{\pm}})}{M_{Z^{\prime}}^2}+\frac{(M_W^2-m^2_{H^{\pm}})^2}{M_{Z^{\prime}}^4}\right]\nonumber\\
&\times&
\sqrt{1-2\frac{(M_W^2+m^2_{H^{\pm}})}{M_{Z^{\prime}}^2}+\frac{(M_W^2-m^2_{H^{\pm}})^2}{M_{Z^{\prime}}^4}}.
\end{eqnarray}
Here $m_{H^{\pm}}$ is the mass of the $H^{\pm}$ Higgs boson and
$M_W$ is the mass of the $W$--boson.

The  $ZW^+W^-$ coupling of the Standard Model will induce, through
$Z-Z^{\prime}$ mixing, a $Z^{\prime}W^+W^-$ coupling. The decay of
$Z^{\prime}$ to a pair of $W^+W^-$ is found to be \cite{Vbarger}
\begin{eqnarray}\label{A464}
\Gamma(Z^{\prime}\rightarrow W^+W^-)=
\frac{g_2^2}{192\pi}\cos^2{\theta_W}\sin^2{\xi}M_{Z^{\prime}}\frac{M_{Z^{\prime}}^4}{M_W^4}
\left(1+20\frac{M_W^2}{M_{Z^{\prime}}^2}+12\frac{M_W^4}{M_{Z^{\prime}}^4}\right)
\left(1-4\frac{M_W^2}{M_{Z^{\prime}}^2}\right)^{\frac{3}{2}}.
\end{eqnarray}

 We now
discuss the decays of $Z^{\prime}\rightarrow
Zh,ZH,Zh^{\prime},ZH^{\prime}$ as well as $Z^{\prime}\rightarrow h
A,\,h^{\prime}A^{\prime}\,\,etc.$. The relevant Lagrangian is
\begin{eqnarray}\label{A328}
\mathcal{L}&=&2g_xM_{Z^{\prime}}\sum_{i=1}^4(Q_{H_d}\cos{\beta}X_{1i}-Q_{H_u}\sin{\beta}X_{2i}){{Z^\prime}^\mu} Z_{\mu}H_i\nonumber\\
&-&g_x\sum_{i=1}^4(Q_{H_d}\sin{\beta}X_{1i}+Q_{H_u}\cos{\beta}X_{2i}){{Z^\prime}^\mu} H_i^0\stackrel{\leftrightarrow}{\partial}_\mu A\nonumber\\
&-&g_x\sum_{i=1}^4(Q_{S_+}\cos{\psi}X_{3i}+Q_{S_-}\sin{\psi}X_{4i}){{Z^\prime}^\mu}
H_i^0\stackrel{\leftrightarrow}{\partial}_\mu A^{\prime},
\end{eqnarray}
where $H_i^0\,\,(=h,\,\,h^{\prime},\,\,\,H,\,\,H^{\prime})$ are
the neutral CP--even Higgs bosons, $m_{H_i}$ are the masses of the
corresponding Higgs boson, $Q_{S_+}\,\,(Q_{S_-})$ is the $U(1)_x$
charge of the $S_+\,\,(S_-)$ field  and $X_{ij}$ are the matrix
elements of the unitary matrix that diagonalizes the CP--even mass
matrix of Eq. (\ref{A12}). The decay rates are then
\begin{eqnarray}\label{A405}
&~& \Gamma(Z^{\prime}\rightarrow
ZH^0_i)=\frac{g_{x}^2}{48\pi}M_{Z^{\prime}}(Q_{H_d}\cos{\beta}X_{1i}-Q_{H_u}\sin{\beta}X_{2i})^2 \times\nonumber\\
&~&\left[1+2\frac{(5M_Z^2-m^2_{H_i})}{M_{Z^{\prime}}^2}+\frac{(M_Z^2-m^2_{H_i})^2}{M_{Z^{\prime}}^4}\right]
\sqrt{1-2\frac{(M_Z^2+m^2_{H_i})}{M_{Z^{\prime}}^2}+\frac{(M_Z^2-m^2_{H_i})^2}{M_{Z^{\prime}}^4}},
\end{eqnarray}
\begin{eqnarray}
\Gamma(Z^{\prime}\rightarrow H_i
A)&=&\frac{g_{x}^2}{48\pi}M_{Z^{\prime}}(Q_{H_d}\sin{\beta}X_{1i}+Q_{H_u}\cos{\beta}X_{2i})^2
\nn\\&\times&\left[1-2\frac{(m_A^2+m^2_{H_i})}{M_{Z^{\prime}}^2}+\frac{(m_A^2-m^2_{H_i})^2}{M_{Z^{\prime}}^4}\right]^{\frac{3}{2}}\\
\Gamma(Z^{\prime}\rightarrow H_i
A^{\prime})&=&\frac{g_{x}^2}{48\pi}M_{Z^{\prime}}(Q_{S_+}\cos{\psi}X_{3i}+Q_{S_-}\sin{\psi}X_{4i})^2
\nn\\&\times&\left[1-2\frac{(m_{A^\prime}^2+m^2_{H_i})}{M_{Z^{\prime}}^2}+\frac{(m_{A^\prime}^2-m^2_{H_i})^2}{M_{Z^{\prime}}^4}\right]^{\frac{3}{2}},
\end{eqnarray}
where $m_A$ and $m_{A^\prime}$ are the pseudoscalar Higgs boson
masses.

 We parameterize the interactions between the neutralinos
($\tilde{\chi}^0_1,\,\tilde{\chi}^0_2,...\tilde{\chi}^0_7$) and
the $Z^{\prime}$ boson as
\begin{eqnarray}\label{A310}
\mathcal{L}&=&\sum_{i,j}g_{ij}\bar{\tilde{\chi}}^0_i\gamma^{\mu}\gamma_5\tilde{\chi}^0_jZ^{\prime}_{\mu}.
\end{eqnarray}
Here the coupling $g_{ij}$ is obtained from the eigenvectors of
the neutralino mass matrix of Eq. (\ref{A16}) as
\begin{eqnarray}\label{A311}
\hat{g}&=&\frac{g_x}{2}O\pmatrix{0&0&0&0&0&0&0\cr 0&0&0&0&0&0&0
\cr0&0&-\frac{x}{2}&0&0&0&0\cr0&0&0&\frac{x}{2}&0&0&0\cr
0&0&0&0&0&0&0\cr 0&0&0&0&0&2&0\cr0&0&0&0&0&0&-2}O^{T},
\end{eqnarray}
with $g_{ij}=(\hat{g})_{ij}$. Here $O$ is the orthogonal matrix
that diagonalizes the neutralino mass matrix. The $Z^{\prime}$
partial decay rates into neutralinos is found to be
\begin{eqnarray}\label{A312}
\Gamma(Z^{\prime}\rightarrow
\tilde{\chi}^0_i\tilde{\chi}^0_i)&=&\frac{g_{ii}^2}{6\pi}M_{Z^{\prime}}\left(1-4\frac{m_i^2}{M_{Z^{\prime}}^2}
\right)^{\frac{3}{2}},\\
\Gamma(Z^{\prime}\rightarrow
\tilde{\chi}^0_i\tilde{\chi}^0_j)&=&\frac{(g_{ij}+g_{ji})^2}{12\pi}M_{Z^{\prime}}\left[1-\frac{(m_i^2+m_j^2)}{2M_{Z^{\prime}}^2}
-\frac{(m_i^2-m_j^2)^2}{2M_{Z^{\prime}}^4}-3\frac{m_im_j}{M_{Z^{\prime}}^2}\right]\nonumber\\
&\times&\sqrt{\left(1-\frac{(m_i+m_j)^2}{M_{Z^{\prime}}^2}\right)\left(1-\frac{(m_i-m_j)^2}{M_{Z^{\prime}}^2}\right)}\qquad\,\,\,\,\,\,(i\neq
j)
\end{eqnarray}
where $m_i$ are the neutralino masses. (Here our result disagrees
with Eq. (48) of Ref. \cite{Gherghetta} by a factor of 2.)

 The Lagrangian for the couplings of $Z^{\prime}$ to the
charginos is given by \cite{Gherghetta}
\begin{eqnarray}\label{A314}
\mathcal{L}&=&\frac{1}{2}g_x\sum_{i,j=1}^2\bar{\tilde{\chi}}_i^{\pm}\gamma^{\mu}(v_{ij}+a_{ij}\gamma_5)\tilde{\chi}_j^\pm
Z^{\prime}_{\mu}.
\end{eqnarray}
The $Z^{\prime}$ decay rate into the chargino pair is then
\begin{eqnarray}\label{A315}
\Gamma(Z^{\prime}\rightarrow
\tilde{\chi}_i^{\pm}\tilde{\chi}_j^{\mp})&=&\frac{g_x^2}{48\pi}M_{Z^{\prime}}\left[(v_{ij}^2+a_{ij}^2)(1-\frac{(m_i^2+m_j^2)}{2M_{Z^{\prime}}^2}
-\frac{(m_i^2-m_j^2)^2}{2M_{Z^{\prime}}^4})+3(v_{ij}^2-a_{ij}^2)\frac{m_im_j}{M_{Z^{\prime}}^2}\right]\nonumber\\
&&\times\sqrt{\left(1-\frac{(m_i+m_j)^2}{M_{Z^{\prime}}^2}\right)\left(1-\frac{(m_i-m_j)^2}{M_{Z^{\prime}}^2}\right)}.
\end{eqnarray}
Here $m_i$ is the chargino mass, $v_{ij}$ and $a_{ij}$ are given
in terms of the charges $Q_{H_u},\,\,Q_{H_d}$ and the matrices $U$
and $V$ which diagonalize the chargino mass matrix Eq.
(\ref{A17}), can be explicitly written as \cite{Gherghetta}
\begin{eqnarray}\label{A316}
v_{11}&=&Q_{H_d}V_{12}^2-Q_{H_u}U_{12}^2,\\
a_{11}&=&Q_{H_d}V_{21}^2+Q_{H_u}U_{21}^2,\\
v_{12}&=&v_{21}=Q_{H_d}V_{12}V_{11}-\delta Q_{H_u}U_{12}U_{11},\\
a_{12}&=&a_{21}=Q_{H_d}V_{12}V_{11}+\delta Q_{H_u}U_{12}U_{11},\\
v_{22}&=&Q_{H_d}V_{11}^2-Q_{H_u}U_{11}^2,\\
a_{22}&=&Q_{H_d}V_{22}^2+Q_{H_u}U_{22}^2,
\end{eqnarray}
where $\delta=sgn(m_{\tilde{\chi}_1^\pm})\times
sgn(m_{\tilde{\chi}_2^\pm})$.

In Table \ref{D4} we present the partial decay rates of
$Z^{\prime}$ to two fermions and to two scalars in Model 1. The
total width of $Z^{\prime}$ is 106 GeV (this ignores three body
decays, which are more suppressed). One sees from Table 7 that the
$Z^{\prime}$ decays dominantly to $q\bar{q}$ with
$Br(Z^{\prime}\rightarrow q\bar{q} )\backsimeq 43.93\%$. On the
other hand, $Br(Z^{\prime}\rightarrow e^+e^-)\backsimeq 1.16\%$ in
this case. Thus this $Z^{\prime}$ is leptophobic. We also see that
$Z^{\prime}\rightarrow \tilde{\chi}_i^0\tilde{\chi}_j^0$ and
$Z^{\prime}\rightarrow \tilde{\chi}_i^\pm\tilde{\chi}_j^\mp$ are
 significant. There are also non--negligible decays into two
Higgs particles, with $Z^{\prime}\rightarrow h^{\prime}A^{\prime}$
being the dominant mode in this class. The decay of $Z^{\prime}$
into sfermions is a new production channel for supersymmetric
particles. Decays into sneutrino pairs is the dominant mode in
this category, with $Br(Z^{\prime}\rightarrow
\tilde{\nu}_{L}\tilde{\nu}_{L})\backsim 7.74\%$. The signature
will be $pp\rightarrow Z^{\prime}\rightarrow
\tilde{\nu}_{Li}\tilde{\nu}_{Li}\rightarrow
\ell_i^-\ell_i^-\tilde{\chi}_1^+\tilde{\chi}_1^+$, where the
sneutrino decays into $\ell_i^-\tilde{\chi}_1^+$, with the
subsequent decay $\tilde{\chi}_1^\pm\rightarrow
\tilde{\chi}_1^0+\pi^\pm$, etc.

In Table \ref{D6} we list the $Z^{\prime}$ partial decay rates in
Model 2. $Br(Z^{\prime}\rightarrow e^+e^-)\backsimeq 1.60\%$ in
this case. Other features are very similar to the case of Model 1
(Table 7).
\newpage
~\vspace{0.45in}

\begin{table}[!h]
\begin{center}\small{
\begin{tabular}{|l|c|}
\hline \rule[1.5mm]{0mm}{0pt}Decay Modes of $Z^{\prime}$ &Width
(GeV)
\\\hline
\hline
\rule[1.5mm]{0mm}{0pt}$Z^{\prime}\rightarrow\{\bar{u}u,\,\bar{c}c,\,\bar{t}t\}$
&$\{4.75,\,4.75,\,4.64\}$\\\hline

\rule[1.5mm]{0mm}{0pt}$Z^{\prime}\rightarrow\bar{d}d\,\,(\bar{s}s,\,\bar{b}b)$
&9.59\\\hline

\rule[1.5mm]{0mm}{0pt}$Z^{\prime}\rightarrow\bar{e}e
(\bar{\mu}\mu,\,\bar{\tau}\tau) $ &1.13\\\hline

\rule[1.5mm]{0mm}{0pt}$Z^{\prime}\rightarrow\nu_{eL}\nu_{eL}\,\,(\nu_{\mu
L } \nu_{\mu L},\,\nu_{\tau L} \nu_{\tau L})$ &0.65\\\hline

\rule[1.5mm]{0mm}{0pt}$Z^{\prime}\rightarrow\nu_{eR}\nu_{eR}\,\,(\nu_{\mu
R} \nu_{\mu R},\,\nu_{\tau R} \nu_{\tau R})$ &4.19\\\hline

\rule[1.5mm]{0mm}{0pt}$Z^{\prime}\rightarrow\bar{\omega}_1\omega_{1}$
&0.50\\\hline

\rule[1.5mm]{0mm}{0pt}$Z^{\prime}\rightarrow\{\tilde{\chi}_1\tilde{\chi}_3,\,\tilde{\chi}_1\tilde{\chi}_4,\,\tilde{\chi}_2\tilde{\chi}_4,\,
\tilde{\chi}_3\tilde{\chi}_4,\,
\tilde{\chi}_3\tilde{\chi}_5,\,\tilde{\chi}_4\tilde{\chi}_5,\,\tilde{\chi}_5\tilde{\chi}_5
,\,\tilde{\chi}_5\tilde{\chi}_6\}$
&$\{0.01,\,0.01,\,0.01,\,3.38,\,0.01,\,0.05,\,3.34,\,5.65\}$\\\hline

\rule[1.5mm]{0mm}{0pt}$Z^{\prime}\rightarrow\{\tilde{\chi}_2^+\tilde{\chi}_2^-,\,\tilde{\chi}_1^+\tilde{\chi}_2^-,\,\tilde{\chi}_1^-\tilde{\chi}_2^+\}$
&$\{3.36,\,0.02,\,0.02\}$\\\hline

\rule[1.5mm]{0mm}{0pt}$Z^{\prime}\rightarrow\tilde{u}_R^{\ast}\tilde{u}_R\,\,(\tilde{c}_R^{\ast}\tilde{c}_R)$
&0.13\\\hline

\rule[1.5mm]{0mm}{0pt}$Z^{\prime}\rightarrow\{\tilde{t}_R^{\ast}\tilde{t}_R,\,\tilde{t}_L^{\ast}\tilde{t}_R,\,\tilde{t}_R^{\ast}\tilde{t}_L\}$
&$\{0.88,\,0.13,\,0.13\}$\\\hline

\rule[1.5mm]{0mm}{0pt}$Z^{\prime}\rightarrow\tilde{e}_L^{\ast}\tilde{e}_L\,\,(\tilde{\mu}_L^{\ast}\tilde{\mu}_L,\,\,\tilde{\tau}_L^{\ast}\tilde{\tau}_L)$
&0.30\\\hline

\rule[1.5mm]{0mm}{0pt}$Z^{\prime}\rightarrow\tilde{e}_R^{\ast}\tilde{e}_R\,\,(\tilde{\mu}_R^{\ast}\tilde{\mu}_R,\,\,\tilde{\tau}_R^{\ast}\tilde{\tau}_R)$
&0.23\\\hline

\rule[1.5mm]{0mm}{0pt}$Z^{\prime}\rightarrow\tilde{\nu}_{eL}^{\ast}\tilde{\nu}_{eL}\,\,(\tilde{\nu}_{\mu
L}^{\ast}\tilde{\nu}_{\mu L},\,\,\tilde{\nu}_{\tau
L}^{\ast}\tilde{\nu}_{\tau L})$ &2.52\\\hline

\rule[1.5mm]{0mm}{0pt}$Z^{\prime}\rightarrow\tilde{\nu}^c_{1s}\tilde{\nu}^c_{1p}\,\,\{\tilde{\nu}^c_{2s}\tilde{\nu}^c_{2p},\,\,\,\tilde{\nu}^c_{3s}\tilde{\nu}^c_{3p}\}$
&1.94\\\hline

\rule[1.5mm]{0mm}{0pt}$Z^{\prime}\rightarrow\tilde{\omega}_{1s}\tilde{\omega}_{1p}$
&0.36\\\hline

\rule[1.5mm]{0mm}{0pt}$Z^{\prime}\rightarrow Zh$ &$1.11$\\\hline

\rule[1.5mm]{0mm}{0pt}$Z^{\prime}\rightarrow \{hA^{\prime},\,HA,\,
h^{\prime}A^{\prime}\}$ &$\{0.03,\,0.47,\,0.62\}$\\\hline

 \rule[1.5mm]{0mm}{0pt}$Z^{\prime}\rightarrow H^+H^-$
&0.46\\\hline

\rule[1.5mm]{0mm}{0pt}$Z^{\prime}\rightarrow W^+W^-$ &1.08\\\hline

\rule[1.5mm]{0mm}{0pt}$Z^{\prime}\rightarrow W^\pm H^\mp$
&0\\\hline
\end{tabular}
\caption{\footnotesize \small{Decay modes for $Z^{\prime}$ in
Model 1 for the parameters used in Table \ref{D3}. The total decay
width is $\Gamma(Z^{\prime}\rightarrow all)=97.68$ GeV.}}
  \label{D4}}
  \end{center}
\end{table}

\newpage

~\vspace{0.4in}

\begin{table}[!h]
\begin{center}\small{
\begin{tabular}{|l|c|}
\hline \rule[1.5mm]{0mm}{0pt}Decay Modes of $Z^{\prime}$ &Width
(GeV)
\\\hline
\hline
\rule[1.5mm]{0mm}{0pt}$Z^{\prime}\rightarrow\{\bar{u}u,\,\bar{c}c,\,\bar{t}t\}$
&$\{15.00,\,15.00,\,14.86\}$
\\\hline
\rule[1.5mm]{0mm}{0pt}$Z^{\prime}\rightarrow\bar{d}d\,\,(\bar{s}s,\,\bar{b}b)$
&20.90\\\hline
\rule[1.5mm]{0mm}{0pt}$Z^{\prime}\rightarrow\bar{e}e
(\bar{\mu}\mu,\,\bar{\tau}\tau) $ &3.69\\\hline

\rule[1.5mm]{0mm}{0pt}$Z^{\prime}\rightarrow\nu_{eL}\nu_{eL}\,\,(\nu_{\mu
L } \nu_{\mu L},\,\nu_{\tau L} \nu_{\tau L})$ &0.37\\\hline

\rule[1.5mm]{0mm}{0pt}$Z^{\prime}\rightarrow\nu_{eR}\nu_{eR}\,\,(\nu_{\mu
R} \nu_{\mu R},\,\nu_{\tau R} \nu_{\tau R})$ &6.19\\\hline

\rule[1.5mm]{0mm}{0pt}$Z^{\prime}\rightarrow\{\bar{\omega}_{1}\omega_{1},\,\bar{\omega}_{1}\omega_{2}\}$
&$\{1.41,\,0.06\}$\\\hline

\rule[1.5mm]{0mm}{0pt}$Z^{\prime}\rightarrow\{\tilde{\chi}_1\tilde{\chi}_3,\,\tilde{\chi}_1\tilde{\chi}_4
,\,\tilde{\chi}_2\tilde{\chi}_4,\,\tilde{\chi}_3\tilde{\chi}_4,
\,\tilde{\chi}_3\tilde{\chi}_5,\,\tilde{\chi}_4\tilde{\chi}_5,\,\tilde{\chi}_5\tilde{\chi}_5,\,\tilde{\chi}_5\tilde{\chi}_6\}$
&$\{0.03\,\,0.03,\,0.03,\,10.99,\,0.01,\,0.04,\,1.63,\,6.64\}$\\\hline

\rule[1.5mm]{0mm}{0pt}$Z^{\prime}\rightarrow\{\tilde{\chi}_2^+\tilde{\chi}_2^-\}$
&$\{10.96\}$\\\hline

\rule[1.5mm]{0mm}{0pt}$Z^{\prime}\rightarrow\tilde{u}^{\ast}_L\tilde{u}_L\,\,(\tilde{c}^{\ast}_L\tilde{c}_L)$
&0.02\\\hline
\rule[1.5mm]{0mm}{0pt}$Z^{\prime}\rightarrow\tilde{u}_R^{\ast}\tilde{u}_R\,\,(\tilde{c}_R^{\ast}\tilde{c}_R)$
&3.80\\\hline
\rule[1.5mm]{0mm}{0pt}$Z^{\prime}\rightarrow\{\tilde{t}_R^{\ast}\tilde{t}_R,\,\tilde{t}_L^{\ast}\tilde{t}_R,\,\tilde{t}_R^{\ast}\tilde{t}_L\}$
&\{5.93,\,0.45,\,0.45\}\\\hline
\rule[1.5mm]{0mm}{0pt}$Z^{\prime}\rightarrow\tilde{d}_L^{\ast}\tilde{d}_L\,\,(\tilde{s}_L^{\ast}\tilde{s}_L,\,\,\tilde{b}_L^{\ast}\tilde{b}_L)$
&0.02\\\hline
\rule[1.5mm]{0mm}{0pt}$Z^{\prime}\rightarrow\tilde{d}_R^{\ast}\tilde{d}_R\,\,(\tilde{s}_R^{\ast}\tilde{s}_R,\,\,\tilde{b}_R^{\ast}\tilde{b}_R)$
&3.77\\\hline

\rule[1.5mm]{0mm}{0pt}$Z^{\prime}\rightarrow\tilde{e}_L^{\ast}\tilde{e}_L\,\,(\tilde{\mu}_L^{\ast}\tilde{\mu}_L,\,\,\tilde{\tau}_L^{\ast}\tilde{\tau}_L)$
&0.18\\\hline
\rule[1.5mm]{0mm}{0pt}$Z^{\prime}\rightarrow\tilde{e}_R^{\ast}\tilde{e}_R\,\,(\tilde{\mu}_R^{\ast}\tilde{\mu}_R,\,\,\tilde{\tau}_R^{\ast}\tilde{\tau}_R)$
&1.54\\\hline
\rule[1.5mm]{0mm}{0pt}$Z^{\prime}\rightarrow\tilde{\nu}_{eL}^{\ast}\tilde{\nu}_{eL}\,\,(\tilde{\nu}_{\mu
L}^{\ast}\tilde{\nu}_{\mu L},\,\,\tilde{\nu}_{\tau
L}^{\ast}\tilde{\nu}_{\tau L})$ &4.54\\\hline
\rule[1.5mm]{0mm}{0pt}$Z^{\prime}\rightarrow\tilde{\nu}^c_{1s}\tilde{\nu}^c_{1p}\,\,\{\tilde{\nu}^c_{2s}\tilde{\nu}^c_{2p},\,\,\,\tilde{\nu}^c_{3s}\tilde{\nu}^c_{3p}\}$
&1.04\\\hline

\rule[1.5mm]{0mm}{0pt}$Z^{\prime}\rightarrow\tilde{\omega}_{1s}\tilde{\omega}_{1p}$
&0.91\\\hline \rule[1.5mm]{0mm}{0pt}$Z^{\prime}\rightarrow Zh$
&$2.96$\\\hline

\rule[1.5mm]{0mm}{0pt}$Z^{\prime}\rightarrow
\{hA^{\prime},\,HA,\,h^{\prime}A^{\prime}\}$
&$\{0.01,\,2.38,\,0.60\}$\\\hline
\rule[1.5mm]{0mm}{0pt}$Z^{\prime}\rightarrow H^+H^-$ &2.38\\\hline
\rule[1.5mm]{0mm}{0pt}$Z^{\prime}\rightarrow W^+W^-$ &2.81\\\hline
\rule[1.5mm]{0mm}{0pt}$Z^{\prime}\rightarrow W^\pm H^\mp$
&0\\\hline
\end{tabular}
\caption{\footnotesize \small{Decay modes for $Z^{\prime}$ in
Model 2 for the parameters used in Table \ref{D5}. The total decay
width is $\Gamma(Z^{\prime}\rightarrow all)=229.93 $ GeV.}}
  \label{D6}}
  \end{center}
\end{table}
\newpage
\section{Other Experimental Signatures}
In this section we discuss experimental signatures of the model
other than $Z^{\prime}$ decays.
\subsection{Z Decay and Precision Electroweak Data}

  The $Z-Z^{\prime}$ mixing angle and the direct coupling of $Z^{\prime}$ to the Standard Model fermions leads to modification of $Z$
  decays. Precision electroweak data from LEP and SLC can be used
  to constrain such a $Z^{\prime}$ in the mass range of a few TeV.
  Typically one finds the $Z-Z^{\prime}$ mixing angle $\xi$
  bounded to be less than a few $\times 10^{-3}$ [4], which is satisfied in our
  models.

The mixing of $Z$ with $Z^{\prime}$ shifts the mass of the $Z$
boson from its SM value, while leaving the $W$ mass unaffected.
This leads to a positive shift in the $\rho$ parameter:
\begin{eqnarray}\label{A.7722}
\rho=\rho_{SM}\left(1+\xi^2\frac{M_{Z^{\prime}}^2}{M_Z^2}\right).
\end{eqnarray}

The partial decay width  $\Gamma (Z\rightarrow f\bar{f})$ is
modified to
\begin{eqnarray}\label{A.773}
\Gamma(Z\rightarrow f\bar{f})=\frac{\alpha M_Z
}{12\sin^2{\theta_{W}}\cos^2{\theta_W}}\left[(g_V\cos{\xi}+\kappa
v_f\sin{\xi})^2+(g_A\cos{\xi}+\kappa a_f\sin{\xi})^2\right].
\end{eqnarray}
where
\begin{eqnarray}\label{A.772}
g_{V}&=&(T_3-2q\sin^2{\theta_W}),\,\,\,\,g_{A}=T_3,\,\,\,\,\kappa=\frac{2g_x\sin{\theta_{W}}\cos{\theta_W}}{e},
\end{eqnarray}
with $q$ being the electric charge of the fermion. $v_f$ and $v_a$
are given in Eqs. (\ref{A401}) and (59).

Partial widths of the $Z$ will deviate from the Standard Model
values owing to the shift in the coupling of $Z$ to fermions as
well as due to a change in the derived value of
$\sin^2{\theta_W}$. We define
\begin{eqnarray}\label{A.7721}\Delta_{f}=\frac{\Gamma (Z\rightarrow f\bar{f})}{\Gamma (Z\rightarrow
f\bar{f})_{SM}}-1.
\end{eqnarray}
We use $\sin^2{\theta^{SM}_W}=0.23113$ (the best fit in the
Standard Model) for evaluating $\Gamma(Z\rightarrow
f\bar{f})_{SM}$. We do not perform a global fit to the available
data, but we present a specific fit which is at least as good as
the Standard Model and perhaps slightly better. We choose to set
$\Delta_{\ell}=0$, which yields $\sin^2{\theta_W}=0.230717$ in
Model 1. With this value of $\sin^2{\theta_W}$ we find
\begin{eqnarray}\label{A.779}
\{\Delta_u,\,\,\Delta_d,\,\,\Delta_{\nu}\}=\{0.00100,\,\,0.00171,\,\,
0.00206\}\,\,\,\,\,\,(\textrm{Model 1}).
\end{eqnarray}
This leads to the following modifications of decay widths:
\begin{eqnarray}\label{A.7791}
\Gamma_{had}&=&\Gamma_{had}^{SM}+\Delta_{d}(2\Gamma_{d}^{SM}+\Gamma_{b}^{SM})+2\Delta_{u}\Gamma_{u}^{SM}=1.74545 \textrm{ GeV},\\
\Gamma_{inv}&=&(1+\Delta_{\nu})\Gamma_{inv}^{SM}=502.793\textrm{ MeV},\\
R_\ell&=&\frac{\Gamma_{had}}{\Gamma(Z\rightarrow
\ell^+\ell^-)}=20.7744.
\end{eqnarray}
We see that $\Gamma_{had}$ is closer to the experimental value of
1.7444 GeV compared to the Standard Model value of 1.7429 GeV.
Similarly $R_{\ell}$ is closer to the experimental value
($20.767\pm 0.025$) than the Standard Model value ($20.744$). On
the other hand, $\Gamma_{inv}$ is somewhat worse than the Standard
Model fit (501.76 MeV) to be compared with the experimental value
of ($499.0\pm 1.5$ MeV). This deviation is still within acceptable
range. Here for our numerical fits we used the central values
$\Gamma^{SM}_{d}=0.383185$ GeV$,\,\,\Gamma^{SM}_{b}=0.375926$ GeV$
$ and $ \Gamma^{SM}_{c}=\Gamma^{SM}_u=0.300302 $ GeV$ $, and
$\Gamma_{had}^{SM} = 1.7429$ GeV \cite{Particle}.

The predicted value of $M_W$ is modified as
\begin{eqnarray}\label{A.781}
M_W&=&\sqrt{\left[\left(1+\xi^2\frac{M_{Z^{\prime}}^2}{M_Z^2}\right)\frac{1-\sin^2{\theta_W}}{1-\sin^2{\theta_W^{SM}}}\right]}M_W^{SM}=80.4427
\textrm{ GeV},
\end{eqnarray}
where $M_W^{SM}=80.391$ GeV is used. This value is closer to the
direct measurement $M_W=80.446$ than the Standard Model value.

In Model 2 we find, following the same procedure,
$\sin^2{\theta_W}=0.230783$, $\Delta_{d}=0.00131$,
$\Delta_{u}=0.00089$, $\Delta_{\nu}=0.00138$
 and
$\Gamma_{had}=1.74493 \textrm{
GeV},\,\,\Gamma_{inv}=502.453\textrm{ MeV},\,\,
R_\ell=20.7682,\,\, M_W=80.4356\textrm{ GeV}. $

The radiative correction parameter in $\mu$ decay, $\Delta r$, is
slightly different in our model compared to the Standard Model. In
the on--shell scheme we have
\begin{eqnarray}
\frac{M_W^2\sin^2{\theta_W}}{(M_W^2\sin^2{\theta_W})_{SM}}=\frac{(1-\Delta
r)_{SM}}{(1-\Delta r)}.
\end{eqnarray}
We obtain $\Delta r=0.03501$ (in Model 1) using the Standard Model
value of $\Delta r=0.0355\pm 0.0019$. Clearly, such a shift is
consistent with experimental constraints ($(\Delta
r)_{exp}=0.0347\pm 0.0011$).
\subsection{$Z^{\prime}$ Mass Limit}
The direct limit on the mass of $Z^{\prime}$ with generic
couplings to quarks and leptons is $M_{Z^{\prime}}>600$ GeV. There
is also a constraint on $M_{Z^{\prime}}$ from the process
  $e^+e^-\rightarrow \mu^+\mu^-$. LEP II has set severe
  constraints on lepton compositeness \cite{Eichten,Particle} from this
  process. We focus on one such amplitude, involving all
  left--handed lepton fields. In our model, the effective
  Lagrangian for this process is
\begin{eqnarray}\label{A.77}
\textit{L}^{\rm eff}
=-g_{x}^2\left(1-\frac{x}{2}\right)^2\frac{1}{M_{Z^{\prime}}^{2}}(\bar{e_{L}}\gamma_{\mu}e_{L})(\bar{\mu_{L}}\gamma^{\mu}
\mu_{L}).
\end{eqnarray}
Comparing with $\Lambda_{LL}^{-}(ee \mu \mu)>6.3$ TeV
\cite{Particle}, we obtain $\frac{M_{Z^{\prime}}}{g_x}\geq
(1-\frac{x}{2})\,2.51$ TeV. For $g_x=0.41 ~(0.45)$ and $x=1.3
~(1.6)$ this implies $M_{Z^{\prime}}\geqslant$ 361 (226) GeV. For
the choice of parameters in Tables \ref{D3} and \ref{D5}, the
above constraint is easily satisfied.

\subsection{$h\rightarrow h^{\prime}h^{\prime}$ Decay}

Since the neutral Higgs boson $h^{\prime}$ is lighter than the
Standard Model Higgs $h$, the decay $h\rightarrow
h^{\prime}h^{\prime}$ can proceed for part of the parameter space.
The decay rate is given by
\begin{eqnarray}\label{F1}
  \Gamma (h\rightarrow h^{\prime}h^{\prime})=\frac{G_{hh^{\prime}}^2}{8\pi
  m_{h}}\sqrt{1-4\frac{m_{h^{\prime}}^2}{m_h^2}},
\end{eqnarray}
where
\begin{eqnarray}\label{F2}
  G^2_{hh^{\prime}}&=&\frac{(g_{1}^2+g_2^2)}{4\sqrt{2}}\left[(\upsilon_{d}{X}_{11}-\upsilon_{u}{X}_{21})({X}_{12}^2-{X}_{22}^2)
  +2(\upsilon_{d}{X}_{12}-\upsilon_{u}{X}_{22})({X}_{11}{X}_{12}-{X}_{21}{X}_{22})\right]\nonumber\\
&+&\frac{g_{x}^2}{4\sqrt{2}}\left[2(4{X}_{31}{X}_{32}-4{X}_{41}{X}_{42}-x{X}_{11}{X}_{12}+x{X}_{21}{X}_{22})\right.\nn\\
&\times &\left.(-x\upsilon_{d}{X}_{12}+x\upsilon_{u}{X}_{22}-4y{X}_{42}+4z{X}_{32})\right.\nonumber\\
&-&\left.(4{X}_{32}^2-4{X}_{42}^2-x{X}_{12}^2+x{X}_{22}^2)
(x\upsilon_{d}{X}_{11}-x\upsilon_{u}{X}_{21}+4y{X}_{41}-4z{X}_{31})\right].
\end{eqnarray}
Here ${X}$ is the unitary matrix that diagonalizes the CP--even
Higgs mass matrix of Eq. (\ref{A12}). In principle this can
compete with the dominant decay $h\rightarrow b\bar{b}$. However
we find that in Model 1 of Table 2 the decay is kinematically
suppressed, while in Model 2 of Table 7 due to the small admixture
of $h$ in $S_+,\,S_-$, this decay is suppressed: $ \Gamma
(h\rightarrow h^{\prime}h^{\prime})=1.48\times 10^{-7}$ GeV (see
Table 11). It is worth noting that if the mixings are as large as
in Table 6 and if the decay is kinematically allowed, then $
\Gamma (h\rightarrow h^{\prime}h^{\prime})\sim 0.1$ MeV is
possible. Once produced, the dominant decays of $h^{\prime}$ will
be $h^{\prime}\rightarrow b\bar{b}$ and $h^{\prime}\rightarrow
c\bar{c}$ with comparable partial widths, as can be seen from
$H_{u}^0$ and $H_d^0$ components in $h^{\prime}$ (see Table 6).

\subsection{Signatures of SUSY Particles}

The supersymmetric particles, once produced in $pp$ ($p\bar{p}$)
collisions, will decay into the LSP. The LSP is $\tilde{\chi}_1^0$
(the neutral Wino) in Model 1 while it is the scalar neutrino
$\tilde{\nu}_L$ in Model 2. In Model 1, $\tilde{\chi}_1^0$ is
nearly mass degenerate with the lightest chargino
$\tilde{\chi}_1^{\pm}$, with a mass splitting of about 180 MeV.
The decay $\tilde{\chi}_1^0\rightarrow \pi^{\pm}\chi_1^{\mp} $
will then occur within the detector. At the Tevetron Run 2 as well
as at the LHC, the process $p\bar{p}$ (or
$pp$)$\rightarrow\tilde{\chi}_1^0+\tilde{\chi}_1^{\pm}$ will
produce these SUSY particles. Naturalness suggest that
$m_{\tilde{\chi}_1^0},\, m_{\tilde{\chi}_1^{\pm}}\lesssim$ 300 GeV
(corresponding to $m_{gluino}\lesssim$ 2 TeV). Strategies for
detecting such a quasi--degenerate pair has been carried out in
Ref. \cite{wells,roy}. In the case where the LSP is the
left--handed sneutrino, the decay $\tilde{\chi}_1^\pm\rightarrow
\ell^{\pm}\tilde{\nu}_L$ will be allowed. In this case
$\tilde{\chi}_1^0$ will decay dominantly to
$\tilde{\chi}_1^0\rightarrow \tilde{\nu}_L\nu_L$.

\section{ Conclusions}
We have suggested in this paper a new class of supersymmetric
$Z^{\prime}$ models motivated by the anomaly mediated
supersymmetry breaking framework. The associated $U(1)$ symmetry
is $U(1)_x=xY-(B-L)$, where $Y$ is the Standard Model hypercharge.
For $1<x<2$, the charges of the lepton doublets and the lepton
singlets have the same sign. This implies that the $U(1)_x$
$D$--term can induce positive masses for both the doublet and the
singlet sleptons and can cure the tachyonic problem of AMSB. We
have shown explicitly that this is indeed possible in this class
of models. In achieving this, the parameters of the model get
essentially fixed. We have found that $M_{Z^{\prime}}=2-4$ TeV and
the $Z-Z^{\prime}$ mixing angle $\xi\simeq 0.001$. The
phenomenologically viable $Z^{\prime}$ turns out to be leptophobic
-- with $Br(Z^{\prime}\rightarrow \ell^+\ell^-)\simeq (1-1.6)\%$.
The dominant decay of $Z^{\prime}$ is to $q\bar{q}$ pair with
$Br(Z^{\prime}\rightarrow q\bar{q})\simeq 44\%$. Decays into
supersymmetric particles and Higgs bosons are also significant.

In Tables \ref{D3} and \ref{D5} we present our spectrum for two
models, Model 1 (with $x=1.3$) and Model 2 (with $x=1.6$). The
lightest SUSY particle is the neutral Wino (Model 1) or the
sneutrino (Model 2). The partial decay widths of $Z^{\prime}$ are
listed in Tables \ref{D4} and \ref{D6}. These models are
compatible with precision electroweak data, with the $Z^{\prime}$
models giving slightly better fits to the data than the Standard
Model. This $Z^{\prime}$ should be within reach of LHC. The
correlations between the $Z^{\prime}$ decays and the
supersymmetric spectrum should make this class of models
distinguishable from other $Z^{\prime}$ models.

\appendix

\section{Appendix}
 In this Appendix we give the one-loop anomalous dimension,
beta-function and the soft SUSY breaking masses for the various
fields in our model.
\subsection{Anomalous  Dimensions}
The one--loop anomalous dimensions for the fields in our model
are:
\begin{eqnarray}\label{A.7}
16\pi^{2}\gamma_{L_{ij}}&=&(Y_{l}Y_l^\dag)_{ji}-\delta_i^j\left(\frac{3}{10}g_{1}^{2}+\frac{3}{2}g_{2}^{2}+2(1-\frac{x}{2})^2g_{x}^{2}\right),\\
16\pi^{2}\gamma_{e^{c}_{ij}}&=&2(Y_l^\dag Y_l)_{ij}-\delta_i^j\left(\frac{6}{5}g_{1}^{2}+2(-1+x)^2g_{x}^{2}\right),\\
16\pi^{2}\gamma_{Q_{ij}}&=&(Y_dY_d^\dag)_{ji}+(Y_uY_u^\dag)_{ji}-\delta_i^j\left(\frac{1}{30}g_{1}^{2}+\frac{3}{2}g_{2}^{2}+\frac{8}{3}g_{3}^{2}+2(\frac{x}{6}-\frac{1}{3})^2g_{x}^2\right),\\
16\pi^{2}\gamma_{U_{ij}}&=&2(Y_{u}^{\dag}Y_{u})_{ij}-\delta_{i}^{j}\left(\frac{8}{15}g_{1}^{2}+\frac{8}{3}g_{3}^{2}+2(\frac{2}{3}x+\frac{1}{3})^2g_{x}^2\right),\\
16\pi^{2}\gamma_{D_{ij}}&=&2(Y_{d}^{\dag}Y_{d})_{ij}-\delta_{i}^{j}\left(\frac{2}{15}g_{1}^{2}+\frac{8}{3}g_{3}^{2}+2(\frac{x}{3}+\frac{1}{3})^2g_{x}^2\right),\\
16\pi^{2}\gamma_{H_{d}}&=&3Y_{d_{3}}^{2}+Y_{l_{3}}^2-\frac{3}{10}g_{1}^{2}-\frac{3}{2}g_{2}^{2}-2\left(-\frac{x}{2}\right)^2g_{x}^2,\\
16\pi^{2}\gamma_{H_{u}}&=&3Y_{u_{3}}^{2}-\frac{3}{10}g_{1}^{2}-\frac{3}{2}g_{2}^{2}-2\left(-\frac{x}{2}\right)^2g_{x}^2,\\
16\pi^{2}\gamma_{\nu^c_{i}}&=&4f_{\nu_{i}^c}^2-2g_{x}^2,\\
16\pi^{2}\gamma_{\nu^c}&=&4f_{\nu^c}^2-2g_{x}^2,\\
16\pi^{2}\gamma_{\bar{\nu}^c}&=&4h^2-2g_{x}^2,\\
16\pi^{2}\gamma_{S_{+}}&=&2 \sum_{i=1}^3f_{\nu_{i}^c}^2+2f_{\nu^c}^2-8g_{x}^2,\\
16\pi^{2}\gamma_{S_{-}}&=&2h^2-8g_{x}^2.
\end{eqnarray}

\subsection{Beta Function }
The beta functions for the Yukawa couplings appearing in the
superpotential, Eq. (4), are:
\begin{eqnarray}\label{A.13}
\beta(Y_{d_{3}})&=&\frac{Y_{d_{3}}}{16\pi^2}\left(6Y_{d_{3}}^2+Y_{u_{3}}^2+Y_{l_{3}}^2
-\frac{7}{15}g_{1}^{2}-3g_{2}^{2}-\frac{16}{3}g_{3}^{2}-\frac{(4+2x+7x^2)}{9}g_{x}^2\right),\\
\beta(Y_{u_{3}})&=&\frac{Y_{u_{3}}}{16\pi^2}\left(6Y_{u_{3}}^2+Y_{d_{3}}^2
-\frac{13}{15}g_{1}^{2}-3g_{2}^{2}-\frac{16}{3}g_{3}^{2}-\frac{(4-10x+13x^2)}{9}g_{x}^2\right),\\
\beta(Y_{l_{3}})&=&\frac{Y_{l_{3}}}{16\pi^2}\left(4Y_{l_{3}}^2+3Y_{d_{3}}^2
-\frac{9}{5}g_{1}^{2}-3g_{2}^{2}-(4-6x+3x^2)g_{x}^2\right),\\
\beta(f_{\nu_{e}})&=&\frac{f_{\nu_{e}}}{16\pi^2}\left(10f_{\nu_{e}}^2+2f_{\nu_{\mu}}^2+2f_{\nu_\tau}^2+2f_{\nu^c}^2
-12g_{x}^2\right),\\
\beta(f_{\nu_{\mu}})&=&\frac{f_{\nu_{\mu}}}{16\pi^2}\left(10f_{\nu_{\mu}}^2+2f_{\nu_{e}}^2+2f_{\nu_\tau}^2+2f_{\nu^c}^2
-12g_{x}^2\right),\\
\beta(f_{\nu_{\tau}})&=&\frac{f_{\nu_{\tau}}}{16\pi^2}\left(10f_{\nu_{\tau}}^2+2f_{\nu_{\mu}}^2+2f_{\nu_e}^2+2f_{\nu^c}^2
-12g_{x}^2\right),\\
\beta(f_{\nu^c})&=&\frac{f_{\nu_{4}}}{16\pi^2}\left(10f_{\nu^c}^2+2f_{\nu_{\mu}}^2+2f_{\nu_\tau}^2+2f_{\nu_e}^2
-12g_{x}^2\right),\\
\beta(h)&=&\frac{h}{16\pi^2}\left(10h -12g_{x}^2\right).
\end{eqnarray}

The gauge beta function of our model are
\begin{eqnarray}\label{A199}
\beta(g_{i})&=&b_{i}\frac{g_{i}^{3}}{16\pi^2},
\end{eqnarray}
where $b_{i}=(\frac{33}{5}, 1, -3, (11x^2-16x+26))$ for
$i=1,\,2,\,2,\,3,\,x$.
\subsection{$A$ terms}
The trilinear soft SUSY breaking terms are given by
\begin{eqnarray}\label{A222}
A_{Y}&=&-\frac{\beta{(Y)}}{Y}M_{aux},
\end{eqnarray}
where $Y=(Y_{u_i},\,
Y_{d_i},\,Y_{l_{i}},\,f_{\nu_{i}^c},\,f_{\nu^c},\,h$).
\subsection{Gaugino  Masses}
The soft masses of the gauginos are given by:
\begin{eqnarray}\label{A.223}
M_{i}&=&\frac{\beta{(g_{i})}}{g_{i}}M_{aux},
\end{eqnarray}
where $i=1,2,3,x$, corresponding to the gauge groups $U(1)_{Y}$,
$SU(2)_{W}$, $SU(3)_{C}$, $U(1)_x$ with $\beta(g_i)$ given as in
Eq. (\ref{A199}) with $M_x=M_1^{\prime}$.

\subsection{Soft SUSY Masses} The soft masses of the squarks
and the sleptons are given in the text. For the $H_u$, $H_d$,
$\nu^c$, $S_+$, $S_-$ fields they are:
\begin{eqnarray}\label{A.291}
(\tilde{m}^{2}_{soft})_{H_{u}}^{H_{u}}&=&\frac{M_{aux}^{2}}{16\pi^2}\left(3Y_{u_{3}}
\beta{(Y_{u_{3}})}-\frac{3}{10}g_{1}\beta{(g_{1})}-\frac{3}{2}g_{2}\beta{(g_{2})}-2\left(\frac{x}{2}\right)^2g_{x}\beta{(g_{x})}\right),\\
(\tilde{m}^{2}_{soft})_{H_{d}}^{H_{d}}&=&\frac{M_{aux}^{2}}{16\pi^2}\left(3Y_{d_{3}}
\beta{(Y_{d_{3}})}+Y_{l_3}\beta{(Y_{l_{3}})}-\frac{3}{10}g_{1}\beta{(g_{1})}-\frac{3}{2}g_{2}\beta{(g_{2})}\right.\nn\\
&-&\left. 2\left(-\frac{x}{2}\right)^2g_{x}\beta{(g_{x})}\right),\\
(\tilde{m}^{2}_{soft})_{S_{+}}^{S_{+}}&=&\frac{M_{aux}^{2}}{16\pi^2}\left(2\sum_{i=1}^3f_{\nu_{i}^c}\beta{(f_{\nu_{i}^c})}+2f_{\nu^c}\beta{(f_{\nu^c})}-8g_{x}\beta{(g_{x})}\right),\\
(\tilde{m}^{2}_{soft})_{S_{-}}^{S_{-}}&=&\frac{M_{aux}^{2}}{16\pi^2}\left(2h\beta{(h)}-8g_{x}\beta{(g_{x})}\right),\\
(\tilde{m}^{2}_{soft})_{\nu_{i}^c}^{\nu_{i}^c}&=&\frac{M_{aux}^{2}}{16\pi^2}\left(4f_{\nu_{i}^c}\beta{(f_{\nu_{i}^c})}-2g_{x}\beta{(g_{x})}\right),\\
(\tilde{m}^{2}_{soft})_{\nu^c}^{\nu^c}&=&\frac{M_{aux}^{2}}{16\pi^2}\left(4f_{\nu^c}\beta{(f_{\nu^c})}-2g_{x}\beta{(g_{x})}\right),\\
(\tilde{m}^{2}_{soft})_{\bar{\nu}^c}^{\bar{\nu}^c}&=&\frac{M_{aux}^{2}}{16\pi^2}\left(4h\beta{(h)}-2g_{x}\beta{(g_{x})}\right).
\end{eqnarray}

\section*{Acknowledgments}
This work is supported in part by DOE Grant \# DE-FG03-98ER-41076
and an award from the Research Corporation. The work of I.G. was
supported in part by the National Science Foundation under grant
PHY00-98791.

\end{document}